\documentclass{article}

\usepackage{arxiv}

\usepackage[utf8]{inputenc} 
\usepackage[T1]{fontenc}    
\usepackage{hyperref}       
\usepackage{url}            
\usepackage{booktabs}       
\usepackage{amsfonts}       
\usepackage{nicefrac}       
\usepackage{microtype}      
\usepackage{lipsum}		
\usepackage{graphicx}
\usepackage{doi}
\usepackage{enumitem}
\usepackage{amsthm}
\theoremstyle{definition}
\newtheorem*{remark}{Remark}

\usepackage{amsmath}

\newcommand{\opA}{\mathop{\vphantom{\sum}\mathchoice
  {\vcenter{\hbox{\huge A}}}
  {\vcenter{\hbox{\Large A}}}{\mathrm{A}}{\mathrm{A}}}\displaylimits}

\hypersetup{
pdftitle={Crosslinking degree variations enable programming and controlling soft fracture via sideways cracking},
pdfsubject={cond-mat.mtrl-sci},
pdfauthor={Miguel Angel Moreno-Mateos, Paul Steinmann},
pdfkeywords={Soft fracture, Elastomers, Phase-field models, Mechanics of materials},
}

\hypersetup{
    colorlinks=true,
    linkcolor=blue,     
    filecolor=magenta,  
    urlcolor=black,      
    citecolor=blue,    
}

\usepackage{booktabs}
\usepackage{amssymb}
\usepackage{amsmath}
\usepackage{stmaryrd}
\usepackage{mathrsfs}
\usepackage{color}
\usepackage{graphicx}
\usepackage{latexsym}
\usepackage{tabls}
\usepackage{graphicx}
\usepackage{epsfig}
\usepackage{subfigure}
\usepackage{rotating}
\usepackage{latexsym}
\usepackage[mathcal]{eucal}
\usepackage{amssymb}
\usepackage{colortbl}
\usepackage{times}
\usepackage{longtable}
\usepackage{algorithmic}
\usepackage{algorithm}
\usepackage{psfrag}
\usepackage{tabularx}
\usepackage{epstopdf}
\usepackage{gensymb}
\usepackage{bbm}
\usepackage{setspace}
\usepackage{fullpage}
\usepackage{url}
\usepackage{calc}
\usepackage{siunitx}

\usepackage{lineno,hyperref}
\usepackage{stmaryrd}
\usepackage{subfigure}
\modulolinenumbers[50]
\usepackage{epstopdf}
\usepackage{array}
\usepackage{textcomp}

\usepackage{booktabs}
\usepackage{mathrsfs}
\usepackage{multirow}
\usepackage{lscape}
\usepackage{amsmath}

\usepackage{mathtools}
\usepackage{amssymb}
\usepackage{color,soul}
\usepackage{bm}

\usepackage{titlesec}
\titleformat{\paragraph}
{\normalfont\normalsize\bfseries}{\theparagraph}{1em}{}
\titlespacing*{\paragraph}
{0pt}{3.25ex plus 1ex minus .2ex}{1.5ex plus .2ex}

\usepackage{tikz}

\usepackage{makecell}
\usepackage{bbm}
\usepackage{float}

\usepackage{lineno}

\usepackage[font=scriptsize]{caption}
\newcommand{\transpose}[1]{#1^{\mathrm{T}}}
\newcommand{\trace}[1]{\mathrm{tr} \, #1}
\newcommand{\cofactor}[1]{\mathrm{cof} \, #1}
\newcommand{\invTrans}[1]{#1^{\mathrm{-T}}}
\newcommand{\Div}[1]{\mathrm{Div} \, #1}

\newcommand{\deformationMap}{\mathbf{\psi}}
\newcommand{\positionMat}{\mathbf{X}}
\newcommand{\positionSpt}{\mathbf{x}}
\newcommand{\detDefGrad}{J}	
\newcommand{\displacement}{\mathbf{u}}
\newcommand{\pressure}{p}
\newcommand{\secondOrderUnitTensorMat}{\mathbf{I}}

\newcommand{\body}{\mathscr{B}}
\newcommand{\bodyMat}{\body_0}
\newcommand{\bodyBoundary}{\partial \body}
\newcommand{\bodyBoundaryMat}{\bodyBoundary_0}
\newcommand{\residualVectorFE}[1]{\mathsf{R}_{#1}}

\newcommand{\volumetric}[1]{\hat{#1}}
\newcommand{\isochoric}[1]{\bar{#1}}

\newcommand{\defGrad}{\mathbf{F}}
\newcommand{\rightCauchyGreen}{\mathbf{C}}
\newcommand{\pkStress}{\mathbf{P}}
\newcommand{\secUnitTensor}{\mathbf{I}}
\newcommand{\strainEnergyFunction}{\Psi}
\newcommand{\interpolantVar}[2]{#1 \thinspace ^{(\mathcal{I}, #2)}} 
\newcommand{\interpolantVarIso}[2]{\isochoric{#1} \thinspace ^{(\mathcal{I}, #2)}} 

\newcommand{\firstInvariant}{\bar{I}_1}
\newcommand{\secondInvariant}{\bar{I}_2}

\newcommand{\controlPoint}[2]{c_{#1}^{(#2)}}
\newcommand{\controlPointVec}[1]{\mathsf{c}^{(#1)}}
\newcommand{\interpPoint}[2]{w_{#1}^{(#2)}}
\newcommand{\interpPointVec}[1]{\mathsf{w}^{(#1)}}
\newcommand{\bspline}[2]{B_{#1}^{(#2)}}
\newcommand{\firstInvariantPts}{n_1}
\newcommand{\secondInvariantPts}{n_2}
\newcommand{\firstInvariantPoint}[1]{\bar{I}_1^{(#1)}}
\newcommand{\secondInvariantPoint}[1]{\bar{I}_2^{(#1)}}

\newcommand{\objectiveFunc}{\mathcal{L}}   
\newcommand{\forceVec}[2]{\mathsf{f}_{#1}^{(#2)}}
\newcommand{\forceWeight}[1]{\omega_f^{(#1)}}
\newcommand{\dispVec}[2]{\mathsf{u}_{#1}^{(#2)}}
\newcommand{\dispWeight}[1]{\omega_u^{(#1)}}

\newcommand{\quantile}[2]{\widehat{Q}_{#1}^{(#2)}}
\newcommand{\quantileWeight}[1]{\lambda_{#1}}
\newcommand{\iCDF}[1]{F_{#1}^{-1}}
\newcommand{\cddf}[1]{F_{#1}}
\newcommand{\pdf}[1]{f_{#1}}

\begin{document}

\title{Biaxial characterization of soft elastomers: experiments and data-adaptive configurational forces for fracture}

\author{ \href{https://orcid.org/0000-0002-3476-2180}{\includegraphics[scale=0.06]{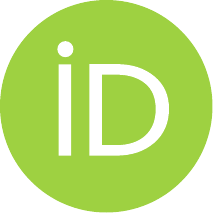}\hspace{1mm}Miguel Angel Moreno-Mateos}\thanks{Corresponding author.} \\
	Institute of Applied Mechanics\\
	Friedrich-Alexander-Universität Erlangen–Nürnberg\\
	91058, Erlangen, Germany\\
	\texttt{miguel.moreno@fau.de} \\
	\And
	\href{https://orcid.org/0009-0008-9625-3446}{\includegraphics[scale=0.06]{orcid.pdf}\hspace{1mm}Simon Wiesheier} \thanks{MAMM and SW contributed equally to this work.}\\
	Institute of Applied Mechanics\\
	Friedrich-Alexander-Universität Erlangen–Nürnberg\\
	91058, Erlangen, Germany\\
	\And
	\href{https://orcid.org/0000-0001-9235-3660}{\includegraphics[scale=0.06]{orcid.pdf}\hspace{1mm}Ali Esmaeili} \\
	Zienkiewicz Institute for Modelling, Data and AI\\
	 Faculty of Science and Engineering, Swansea University\\
	SA1 8EN, Swansea, United Kingdom\\
	\And
	\href{https://orcid.org/0000-0002-4616-1104}{\includegraphics[scale=0.06]{orcid.pdf}\hspace{1mm}Mokarram Hossain} \\
	Zienkiewicz Institute for Modelling, Data and AI\\
	 Faculty of Science and Engineering, Swansea University\\
	SA1 8EN, Swansea, United Kingdom\\
	\And
	\href{https://orcid.org/0000-0003-1490-947X}{\includegraphics[scale=0.06]{orcid.pdf}\hspace{1mm}Paul Steinmann} \\
	Institute of Applied Mechanics\\
	Friedrich-Alexander-Universität Erlangen–Nürnberg\\
	91058, Erlangen, Germany\\
	Glasgow Computational Engineering Centre\\
	University of Glasgow\\
	G12 8QQ, United Kingdom\\
}

\maketitle

\begin{abstract}
Understanding the fracture mechanics of soft solids remains a fundamental challenge due to their complex, nonlinear responses under large deformations. While multiaxial loading is key to probing their mechanical behavior, the role of such loading in fracture processes is still poorly understood. Here, we present a combined experimental–computational framework to investigate fracture in soft elastomers under equi-biaxial loading. We report original equi-biaxial quasi-static experiments on five elastomeric materials, revealing a spectrum of material and fracture behavior---from brittle-like to highly deformable response with crack tip strains exceeding \qty{150}{\%}. Motivated by these observations, we develop a hybrid computational testbed that mirrors the experimental setup and enables virtual biaxial tests. Central to this framework are two components: a data-adaptive formulation of hyperelastic energy functions that flexibly captures material behavior, and a post-processing implementation of the Configurational Force Method, providing a computationally efficient estimate of the $J$-integral at the crack tip. Our data-adaptive framework for hyperelastic energy functions proves versatility to capture with high accuracy the hyperelastic behavior observed in the biaxial experiments. This is important because accurately capturing the constitutive behaviour of soft solids is key for a reliable application of the Configurational Force Method to soft solids. In the limit of crack onset, a critical value of the crack tip configurational force allows for a criterion of fracture toughness. Together, our experimental, theoretical, and computational contributions offer a new paradigm for characterizing and designing soft materials with tailored fracture properties.
\end{abstract}


\section{Introduction}
Soft materials such as elastomers and hydrogels exhibit complex material response at large deformations. These properties pose significant challenges for constitutive modeling, particularly under multiaxial loading conditions. Comprehensive characterization of these behaviors is critical, given the widespread use of soft materials in biomedical devices, soft robotics, and flexible electronics \cite{Zhao2017,Moreno-Mateos2022c,Arif2024}. Despite their relevance, experimental studies under biaxial loading remain relatively limited. Seminal contributions include the early work by Treloar \cite{Treloar1944} and later investigations by James and Green \cite{James1975}, Kawabata et al. \cite{Kawabata1981}, and more recent the study by Pancheri and Dorfmann \cite{Pancheri2014}. These efforts provide valuable insights into the complex mechanical response of soft materials under biaxial stretching. However, comprehensive experimental datasets are limited, and studies extending beyond basic material characterization---such as those investigating the fracture behavior of soft materials---remain exceptionally rare \cite{Ahmad2019b,Khiem2019,Esmaeili2023}.

\subsection{Fracture \& configurational mechanics}
Classical fracture mechanics approaches include the energy release rate \cite{Griffith1921}, stress intensity factors \cite{Irwin1957}, the $J$-integral \cite{Cherepanov1967,Rice1968,Shih1986}, and the crack tip opening displacement \cite{Wells1961}. For mixed-mode crack propagation, various deflection criteria have been proposed, such as the maximum tangential stress criterion \cite{Erdogan1963}, the strain energy density $S$-factor \cite{Sih1974}, the maximum energy release rate criterion \cite{Hussain1974}, and the apparent crack extension force approach \cite{Strifors1974}. These established methods are primarily grounded in Linear Elastic Fracture Mechanics (LEFM). At finite strains, the crack tip often exhibits pronounced blunting \cite{Hui2003,Qi2019,Lu2021}, rendering many LEFM assumptions inapplicable. Only a limited number of closed-form solutions exist that analytically describe the deformed crack contour and the near-tip fields in soft materials \cite{Long2011,Long2015}.

Configurational mechanics offers a theoretical framework to quantify the driving forces that produce changes in the material configuration due to the evolution of defects \cite{Eshelby1951,Steinmann2000,Podio-Guidugli2002,Gross2003,Steinmann2008,Steinmann2009,Steinmann2022}. These changes are consistent with a principle of maximal energy dissipation. The energy released during such configurational changes can be described in terms of configurational forces, which are power or variationally conjugated to variations in the material configuration. As such, configurational forces act on the material manifold and drive changes in the material positions.

Computational fracture mechanics offers a range of modeling strategies, many of which are rooted in the energy-based variational approach introduced by Francfort and Marigo \cite{Francfort1998}. Among others, we mention: non-local damage models with internal history variables and non-local strain measures \cite{Schreyer1990,Peerlings1998,Jirasek1998}; element deletion method \cite{Song2008}; cohesive zone models with traction-separation laws \cite{Dugdale1960}; extended finite element method (XFEM) \cite{Moes1999}; mesh-free methods \cite{Belytschko1995}; phase-field methods model fracture \cite{Miehe2010,Kumar2020a,Lo2022,Moreno-Mateos2023,Moreno-Mateos2024n,Dammass2024t,Hu2025}; eigenfracture \cite{Schmidt2009,Storm2023}; peridynamics \cite{Silling2000,Javili2021}. 

Configurational forces\footnote{Configurational forces are also referred to as material forces.} find their most widespread application in the field of fracture mechanics. The contour integral form of the celebrated $J$-integral (cf., e.g., \cite{Shih1986}) is mathematically equivalent to the configurational force at the crack tip. \textit{Pacman-shaped} domains enclosing the crack tip have been used to prove this equivalency \cite{Moreno-Mateos2024b}. The Configurational Force Method identifies the crack tip configurational force---computed as nodal forces resulting from a finite element discretization of the Eshelby stress tensor---as a direct and effective estimator of the $J$-integral \cite{Steinmann2001}. In this context, several authors have proposed crack propagation laws grounded in configurational mechanics \cite{Steinmann2001,Kienzler2002b,Denzer2003,Gurtin1996,Zhou2022}, including derivations for curved crack paths \cite{Schutte2009,Frankl2022,Schmitz2023} and strain gradient elasticity formulations \cite{Serrao2025b}. Other contributions include configurational criteria for mixed-mode fracture \cite{Guo2017} and extensions to fatigue failure \cite{Liu2020s,Yan2023s}. Inelastic fracture behavior has been explored in the context of plasticity and creep \cite{Nguyen2005,Naser2007,Kaliske2009,Ozenc2014,Kolednik2022}. Coupled-field problems have also benefited from this approach, including electro-viscoelastic fracture \cite{Denzer2014}, fracture of dielectric electro-active elastomers \cite{Moreno-Mateos2024a}, and flexoelectric materials \cite{Serrao2025a}. A recent development is configurational peridynamics \cite{Steinmann2023}. These examples illustrate the breadth of applications where configurational mechanics provides a powerful and unified perspective on fracture phenomena.

Interestingly, the advancement in configurational mechanics has shown that configurational forces can serve as a standalone and sufficient framework to describe fracture mechanics of soft materials at finite strains \cite{Moreno-Mateos2024b}. Calculating the configuration forces at the tip of the crack must yield a value equal to the $J$-integral, eliminating the need for analytical solutions. The results may be as accurate as the constitutive model utilized to solve the boundary value problem.

\newpage

\subsection{Data-adaptive constitutive modeling}
Accurate constitutive modeling, essential for capturing material behavior, is necessary to compute reliable configurational forces. Traditionally, material model calibration is based on phenomenological models whose parameters are fitted to experimental data. A wide variety of phenomenological models exist; we refer to \cite{steinmann_hyperelastic_2012, ricker_systematic_2023} for recent state-of-the-art reviews. In recent years, data-driven constitutive modeling approaches have gained increasing attention, as they alleviate the need to predefine a specific model form and move toward model-free or model-discovery paradigms \cite{Dal2023,Tikenogullari2023}. Notable among these is the work by \cite{kirchdoerfer_data-driven_2016}, where each quadrature point is assigned the nearest state from the data set. As the field has matured, multiple research streams have emerged. These range from symbolic regression \cite{abdusalamov_automatic_2023, bahmani_physicsconstrained_2024} and Gaussian Process Regression \cite{frankel_tensor_2019,Ellmer2024} to various neural network (NN)-based approaches \cite{Holthusen2024}.

Several NN formulations have recently been proposed. Physically-Informed Neural Networks (PINNs) \cite{hamel_calibrating_2023} are a promising tool for solving inverse problems. They aim to replace the calibration process with a neural network that simultaneously identifies material and network parameters, but they typically assume the underlying material model is known. In contrast, physics-augmented (PANN) \cite{dammas_neural_2024, klein_neural_2025,Ortigosa2025} and constitutive artificial neural networks (CANN) \cite{linka_constitutive_2021, linka_new_2023, linden_neural_2023} are designed to discover new constitutive models directly from data. In all such approaches, the incorporation of physical knowledge into the training process is critical for identifying plausible and physically consistent models. We refer to \cite{fuhg_review_2025, romer_reduced_2024} for comprehensive reviews on data-driven constitutive laws and calibration techniques.

Experimental data for model calibration are often obtained from simple load cases such as uniaxial or simple shear, which typically result in homogeneous deformation fields. While such experiments are easy to conduct, the resulting data often lack the richness needed to calibrate versatile material models. An alternative is to use more complex geometries and loading conditions that produce inhomogeneous deformation fields. Digital Image Correlation (DIC) \cite{sutton_image_2009} enables the extraction of full-field displacement measurements from these experiments. Together with global force data, this provides a rich experimental dataset for calibration. Common parameter identification approaches are Finite Element Model Updating (FEMU) and the Virtual Fields Method (VFM) \cite{pierron_virtual_2012}. FEMU involves iteratively solving the boundary value problem using the Finite Element Method to replicate the experimental setup. Its main strength lies in the ability to leverage well-established FEM and optimization algorithms. The Efficient Unsupervised Constitutive Law Identification and Discovery (EUCLID) framework works similarly to the VFM where displacement fields are leveraged to construct the stress field. EUCLID works by prescribing a set of canonical material models and filtering the dominant material models using sparse regression, leaving a constitutive law described by only a sparse set of the prescribed material models \cite{flaschel_automated_2023}. Extensions of the original approach, including Bayesian interference (Bayesian-EUCLID) \cite{joshi_bayesian-euclid_2022} and neural networks (NN-EUCLID) \cite{thakolkaran_nn-euclid_2022} have already been developed.

\newpage

\subsection{Our approach: data-adaptive configurational forces}
Despite the body of work in the experimental mechanics of soft solids, data-driven model discovery, and theoretical and computational fracture mechanics, several important gaps in knowledge remain, to the best of the authors’ knowledge. These are: i) experimental data on constitutive behavior of soft materials under biaxial conditions are limited; ii) experimental data on fracture behavior of soft materials in biaxial loading settings are similarly scarce, if not inexistent; iii) data-driven modeling approaches have been only marginally applied to biaxial experimental data; iv) the Configurational Force Method---as an estimator of the $J$-integral---has not yet been employed in combination with data-adaptive constitutive models, nor has it been applied to soft elastomers under biaxial loading conditions. Furthermore, the intersection of these four points is \textit{terra incognita}.

Here, we present a comprehensive study that integrates experiments with a novel computational framework for data-adaptive configurational forces in fracture mechanics. 
On the one hand, we report a comprehensive experimental characterization of soft elastomers under equi-biaxial, quasi-static loading conditions. We provide the results in the main manuscript and also the complete dataset in a Zenodo repository. The experimental insights into the a) constitutive and b) fracture behavior span from brittle constitutive/fracture responses to remarkably soft ones, i.e., highly non-linear at strains above \qty{150}{\%} at the crack tip, depending on the material. The experimental observations are then utilized to motivate a hybrid computational framework able to inform and further expound the experiments---in terms of constitutive behavior \textit{ante} fracture onset and fracture toughness. The framework is built upon two core components: (i) a data-adaptive formulation of hyperelastic energy functions, and (ii) the Configurational Force Method implemented as a post-processing algorithm. 

We propose a versatile, data-adaptive framework for calibrating hyperelastic energy functions by combining FEMU with a novel data-adaptive constitutive modeling approach. In our framework, constructed in the sequel of a previous work of the authors in \cite{Wiesheier2024}, the strain energy function is represented via B-spline interpolation defined over the space of invariants. Unlike phenomenological models, this approach makes no assumptions about the shape of the energy function. Instead, its form is determined by a set of interpolation points (defined in the invariant space) and the corresponding values, both identified automatically using robust non-linear optimization. In the present work, the framework utilizes original experimental data on five different soft elastomers under equi-biaxial conditions consisting of global reaction forces and full-field displacements obtained through DIC. In turn, the Configurational Force Method is implemented as postprocessing algorithm that calculates the Eshelby stress tensor and its corresponding nodal forces, i.e., configurational forces. The total configurational force at the crack tip estimates the $J$-integral. This approach offers a simpler alternative compared to traditional methods that rely on evaluating a contour integral around the crack tip. In the sequel of a previous work of the authors \cite{Moreno-Mateos2024b}, here the Configurational Force Method enables fracture assessment in soft materials under biaxial loading conditions. Overall, by integrating rich biaxial data sets with a flexible, data-adaptive constitutive model, our approach offers a powerful tool to accurately capture complex material behavior. This is paramount because accurately capturing the constitutive behavior of soft elastomers is key for a reliable application of the Configurational Force Method to soft solids.

\section{Experimental characterization of soft elastomers under biaxial loading}
A comprehensive experimental characterization of soft elastomers offers novel insights into the constitutive behavior (\textit{ante} fracture onset) and the fracture behavior under equi-biaxial conditions. Furthermore, the material characterization is central in this work as it will allow for accurate constitutive models and determination of configurational forces for fracture assessment. 

The results are detailed in the main manuscript and the dataset is available in the following Zenodo repository:

• \url{https://doi.org/10.5281/zenodo.15187640}.

\subsection{Materials, synthesis and experimental setup}

We prepared five soft elastomers with stiffnesses ranging from \qty{5}{\kilo\pascal} to \qty{350}{\kilo\pascal}. Four of them were synthesized by curing two-component blends at elevated temperatures: Elastosil P7670 for three different volume mixing ratios, and Sylgard 184. The fifth material is VHB Tape, which is supplied by the manufacturer as pre-formed sheets that can be cut to size. Additional details about the materials are provided below.
\begin{itemize}[itemsep=0pt, parsep=0pt, left=0pt, align=parleft]
\item Elastosil P7670 2:1 (softest), 8:5 (medium), and 1:1 (stiffest) (Wacker Chemie AG, Munich, Bavaria, Germany) was synthesized mixing two raw phases (phase A and B) to form a crosslinked silicone elastomer. The mixing ratio of the phases is defined as the volume ratio between the phases $V_\text{A}/V_\text{B}$, with $V_\text{A}$ and $V_\text{B}$ the volume of the phase A and B, respectively. The three polymer variants are 2:1, 8:5, and 1:1, respectively from the softest to the stiffest (cf. the previous work of the authors \cite{Moreno-Mateos2024n} for additional variants of the elastomer). Each mixture was cast into an open polytetrafluoroethylene (PTFE) mold, degassed under vacuum for \qty{15}{\minute}, and cured in an oven at \qty{120}{\celsius} for \qty{2}{\hour}. The resulting square samples had a uniform thickness of \qty{2}{\milli\meter}.

\item Sylgard 184 (Dow Inc., Midland, Michigan, United States) was prepared mixing two raw phases in a 10:1 volume ratio, resulting in a crosslinked material stiffer than all Elastosil P7670 variants. As with the Elastosil samples, the mixture was cast into an open mold, degassed under vacuum for \qty{15}{\minute}, and cured in an oven at \qty{120}{\celsius}. The final cured sample also had a thickness of \qty{2}{\milli\meter}.

\item VHB 4905 Tape (3M, Saint Paul, Minnesota, United States) was supplied by the manufacturer as pre-formed sheets of thickness $t=\qty{0.5}{\milli\meter}$, which were cut to the desired dimensions. Among all tested elastomers, VHB exhibited the most pronounced viscoelastic behavior. To minimize rate-dependent inelastic effects, all tests involving VHB were conducted at low strain rates ($\qty{0.01}{\per \second}$).
\end{itemize}

\begin{figure}[H]
\centering
\includegraphics[width=0.9\textwidth]{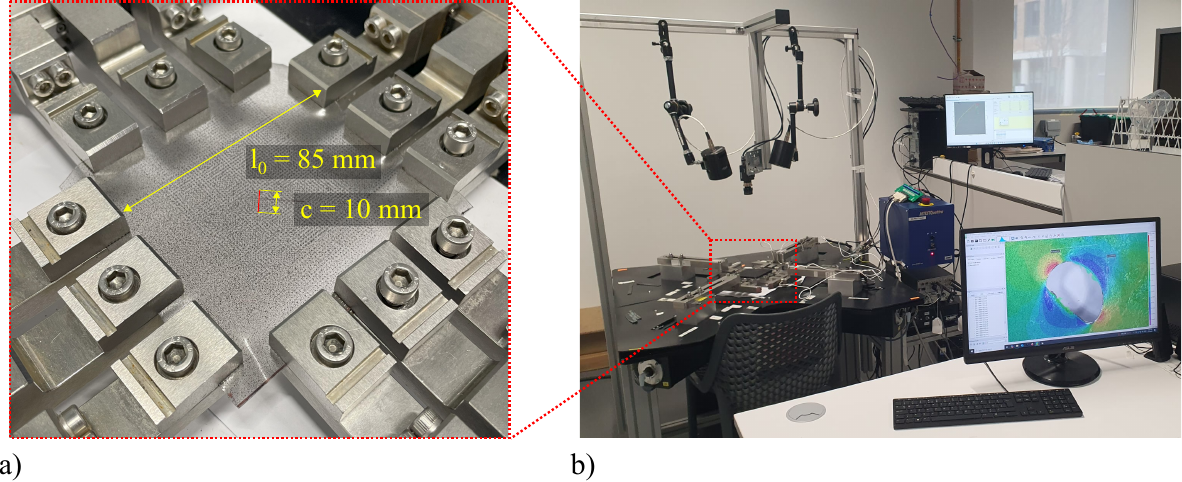}
\caption{\textbf{Experimental setup for biaxial stretch tests.} (a) Close-up view of the grips used in the experimental testbed. Each grip consists of three independent clamps designed to hold rectangular specimens. A sliding grip system allows the gripping fingers to translate perpendicular to the loading direction, i.e., in the direction of the edge of the sample. To that end, the gripping fingers can freely move in this direction. Insets: the initial configuration shows the separation between opposing grips and a centrally positioned notch in the specimen, inclined at an angle of \qty{45}{\degree} relative to the loading axis. This orientation ensures symmetric deformation with respect to the axis of the pre-cut. (b) Overview of the experimental setup, featuring the DIC imaging system mounted above the specimen. The testbed includes two orthogonal pairs of grips that enable biaxial tensile loading. The setup is connected to control and acquisition computers used to operate the system and record images throughout the deformation process.}
\label{fig:Experimental_setup}
\end{figure}

The experimental testbed consisted of a bespoke biaxial testing machine and an imaging system. The biaxial machine (eXpert 8000, ADMET, Norwood, MA, United States) is equipped with a lateral sliding pinching grip system allowing the gripping fingers positioned at the corners to translate perpendicular to the loading direction, i.e., in the direction of the edge of the sample, as the material expanded (see Figure~\ref{fig:Experimental_setup}.a). The initial distance between clamps was $l_0=\qty{85}{\milli \meter}$ for all tests. Equi-biaxial experiments were carried out up to complete rupture on samples featuring an initial notch  of \qty{10}{\milli \meter} centered and oriented at \qty{45}{\degree} with respect to the loading axes of the testing machine (see Figure~\ref{fig:Experimental_setup}.a). A quasi-static strain rate of \qty{0.01}{\per \second}, i.e., displacement rate of the axes of \qty{0.85}{\milli \meter \per \second} was used for all tests. 

The biaxial machine was equipped with a 5.0-megapixel CCD digital camera (2448 x 2028 @ 75 fps) and a 2D DIC System analysis software (VIC-2D, Correlated Solutions, Inc.). DIC was used to compute the displacement fields, which served as input for calibrating data-adaptive hyperelastic strain energy functions—an essential step for accurately computing fracture configurational forces, as detailed in subsequent sections. The camera was mounted perpendicular to the surface of the sample, providing a fixed length-to-pixel scale of \qty{0.0289309}{\milli\meter\per\text{pixel}}. The surface of the samples were sprayed with black paint to generate a speckle pattern. For postprocessing the displacement fields, a subset of \qty{29}{\text{pixel}} and a step size of \qty{7}{\text{pixel}} were used\footnote{For accurate DIC measurements, the subset size is typically recommended to be at least three times larger than the step size.}. In addition, engineering strain fields were derived from the displacement data by computing the Lagrange strain tensor, $\mathbf{E}=\frac{1}{2}\left[ \mathbf{C}-\mathbf{I}\right]$, with $\mathbf{C}$ the right Cauchy-Green tensor and $\mathbf{I}$ the second order identity tensor, making the results insensitive to arbitrary rigid-body motion. Note that the strain fields will not be necessary to calibrate the data-adaptive strain energy functions, which work only on displacement fields. However, we present the strain fields as they provide valuable insight for interpreting the constitutive behavior near the crack tip and for comparing the mechanical response of the different materials.

\subsection{Results}
\begin{figure}[H]
\centering
\includegraphics[width=0.95\textwidth]{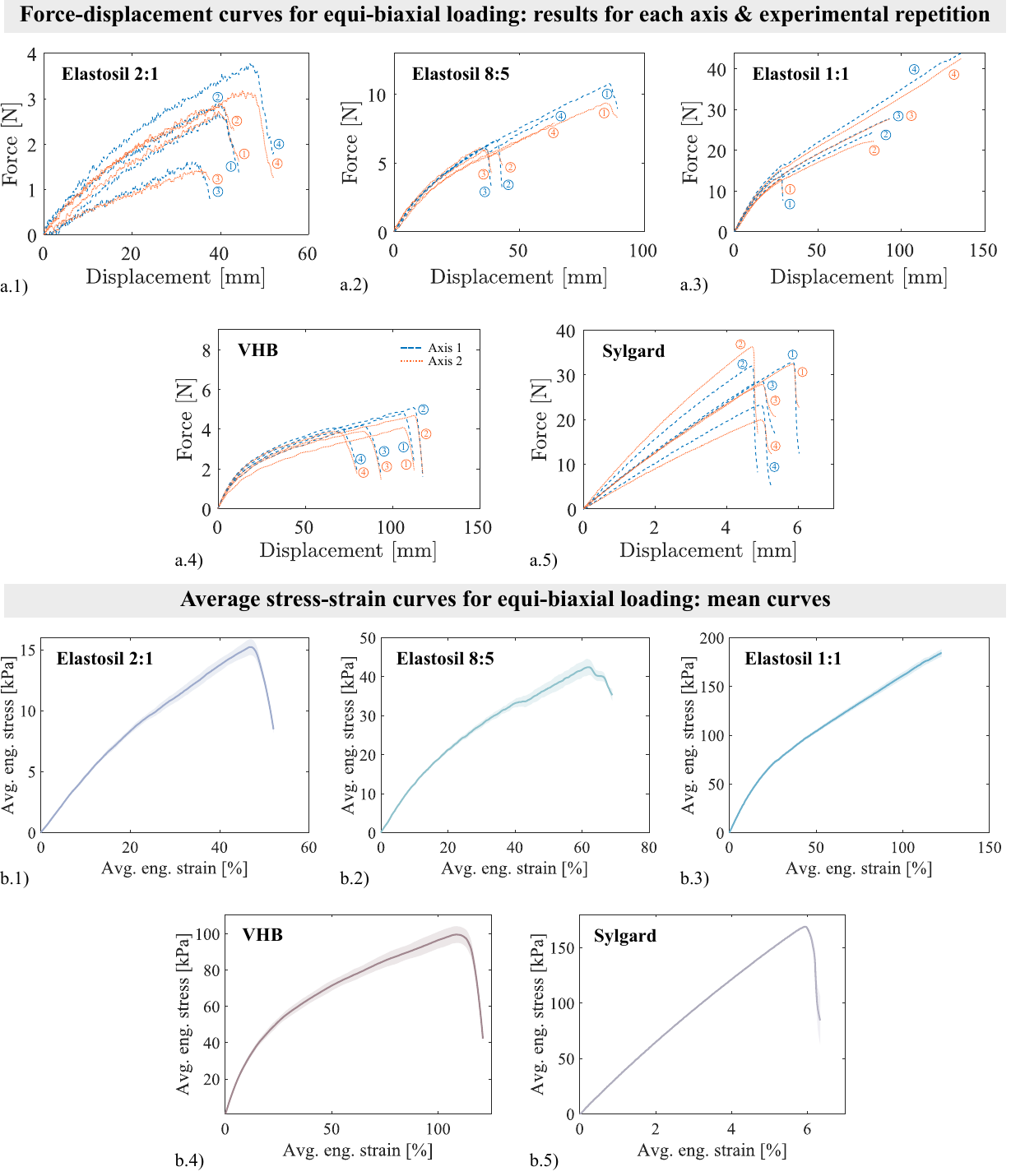}
\caption{\textbf{Experimental results for the biaxial experiments.} The two independent axes move identically at a displacement rate of \qty{0.85}{\milli \meter \per \second}, which renders a strain rate of \qty{0.01}{\per \second}. (a.1-5) Force displacement results for biaxial experiments and for each of the axes of the machine (blue and orange colored-lines). The results for each experimental repetition are marked with \textcircled{\tiny 1}, \textcircled{\tiny 2}, \textcircled{\tiny 3}, and \textcircled{\tiny 4} so that they can be easily identified in the Zenodo dataset.
(b.1-5) Average engineering stress versus average engineering strain curves for equi-biaxial tensile tests on VHB, Sylgard, and Elastosil with mixing ratios of 2:1, 8:5, and 1:1, respectively. The average engineering strain is obtained dividing the value of displacement of the clamps by the initial distance between clamps ($l_\text{0}$). The engineering stress is calculated for each axis as force per unit of initial cross-section area. Scatter areas are included to quantify the variation between the measurements of the two independent axes. Four repetitions are performed for the same tests conditions.}
\label{fig:Force-Disp}
\end{figure}

The experimental characterization highlights a range of fracture behaviors, from brittle-like failure in Sylgard to highly non-linear fracture responses in VHB Tape. The corresponding force-displacement results are shown in Figure~\ref{fig:Force-Disp}. Panel (a) presents the force-displacement curves along each axis for all experimental repetitions, while panel (b) displays the average engineering stress–strain curves. Individual repetitions, labeled 1 through 4, are shown explicitly to facilitate comparison with the raw data—force-displacement measurements and subsequent DIC fields—available on Zenodo. This also enables an assessment of the variability across experiments. The average stress–strain curves in (b), together with the associated scatter for each loading direction, provide a more intuitive basis for interpreting the material response and the influence of experimental variability.


\newpage

\begin{figure}[H]
\centering
\includegraphics[width=0.95\textwidth]{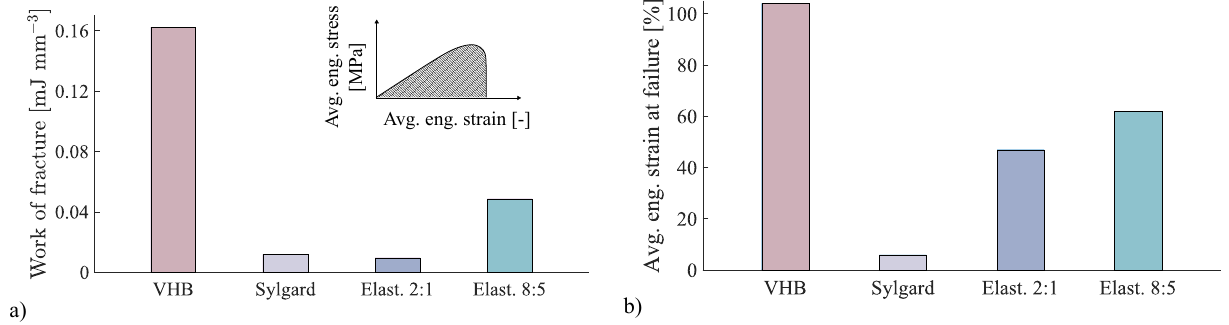}
\caption{\textbf{Experimental total work of fracture and strain at failure for the biaxial experiments.} (a) Work of fracture for each material calculated as the total work performed by both axes during the deformation of the sample until full propagation of the crack divided by the initial volume of the sample. The volume of the samples is \qty{85}{\milli \meter} $\times$ \qty{85}{\milli \meter} $\times$ \qty{2}{\milli \meter} (thickness) for all samples except for VHB. For VHB, thickness is \qty{0.5}{\milli \meter}. Schematic inset: outlines the area under the stress-strain curve used to compute the total work of fracture. 
(b) Average strain at failure for each elastomer calculated as the displacement for the maximum force. The average engineering strain is obtained dividing the value of displacement of the clamps by the initial distance between clamps ($l_\text{0}$). The values are taken as the displacement for the maximum value of stress of the mean curves.
Note that the work of fracture and displacement at failure for Elastosil 1:1 is not included as the samples do not rupture.}
\label{fig:barplots_exp}
\end{figure}

Key metrics derived from the experiments include the total work of fracture and the average strain at failure. Figure~\ref{fig:barplots_exp} presents bar plots of the total work of fracture per unit volume\footnote{In the literature, the total work of fracture is often decomposed into essential and non-essential components. The essential part corresponds to the energy dissipated in the immediate vicinity of the crack tip, i.e., the fracture process zone—and is considered an intrinsic material property. The non-essential part accounts for energy dissipated in the surrounding bulk material. This study does not attempt to distinguish between these two contributions.} for all tested materials, except Elastosil 1:1\footnote{Elastosil 1:1 did not reach complete rupture due to its tendency to fracture laterally. Instead, samples slipped out of the clamps, a behavior attributed to the extreme stretchability enabled by sideways fracture paths (see \cite{Moreno-Mateos2024n}).}. Among all materials, VHB Tape exhibits the highest fracture energy at \qty{0.1624}{\milli \joule \per \milli \meter \cubed}, which is \qty{1376}{\%} higher than Sylgard, \qty{1709}{\%} higher than Elastosil 2:1, and \qty{335}{\%} higher than Elastosil 8:5. In contrast, Sylgard, Elastosil 2:1, and Elastosil 8:5 exhibit fracture energy values of \qty{0.0118}{\milli \joule \per \milli \meter \cubed}, \qty{0.0095}{\milli \joule \per \milli \meter \cubed}, and \qty{0.0485}{\milli \joule \per \milli \meter \cubed}, respectively. Notably, despite being synthesized from the same base components, Elastosil 8:5 shows a markedly higher work of fracture than Elastosil 2:1—a difference of \qty{511}{\%}---highlighting the strong influence of mixing ratio on fracture behavior. The average strain at failure---computed by dividing the clamp displacement by the initial grip separation ($l_0$)\footnote{The average strain at failure provides an average, global measure that does not capture the local strain concentration near the crack tip but that aims at describing the homogeneous structural strain in the direction of the axes.}---follows a trend consistent with the work of fracture. VHB again shows the highest value (\qty{104}{\%}), while Sylgard fails at the lowest strain (\qty{5.9}{\%}). Elastosil 2:1 and 8:5 fall in between, with average failure strains of \qty{46.9}{\%} and \qty{61.8}{\%}, respectively. However, in this case, the increase for Elastosil 8:5 relative to Elastosil 2:1 is more modest—only \qty{132}{\%}. This suggests that while the increased crosslink density in Elastosil 8:5 significantly raises the energy dissipation capacity, it does not proportionally extend the stretchability before rupture.

A critical aspect of the analysis concerns the onset of crack propagation and the corresponding displacement at which it occurs. Distinguishing between the pre-crack and post-crack phases during equi-biaxial deformation is essential: the constitutive models introduced in the following section are calibrated exclusively using data from the pre-crack regime, and the configurational forces are evaluated precisely at the loading stage where crack propagation initiates. To this end, Table~\ref{tab:displacement_crack_onset} reports the displacement and strain values at crack onset for each material and for all experimental repetitions. As in previous figures, each repetition is labeled 1 through 4 to facilitate correlation with the raw dataset available on Zenodo. The results are consistent with the trends observed in the strain at failure: VHB exhibits the highest strain at crack onset, reaching \qty{57.5}{\%}, while Sylgard shows the lowest value at \qty{6.6}{\%}. Elastosil 2:1 and Elastosil 8:5 fall in between, with onset strains of \qty{32.4}{\%} and \qty{35.1}{\%}, respectively. The material characterization plays a central role in this work, since accurately capturing the constitutive behavior of soft elastomers from experimental data is essential for the reliable application of the Configurational Force Method to soft solids.

The fracture typology also varies across materials. VHB, Sylgard, and Elastosil 2:1 and Elastosil 8:5 exhibit forward fracture while Elastosil 1:1 undergoes sideways fracture. As previously reported by the authors \cite{Moreno-Mateos2024n}, high degrees of crosslinking lead to deformation-induced fracture anisotropy, whereby cracks tend to propagate parallel to the direction of maximum stretch. Supplementary Video 1 illustrates this mechanism for one of the Elastosil 1:1 experiments---an elastomer characterized by high crosslink density and sideways fracture behavior. Interestingly, Elastosil 1:1 fails at a lower strain than its less crosslinked counterparts, with a crack onset strain of only \qty{30.8}{\%} (see Table~\ref{tab:displacement_crack_onset}). This indicates that sideways cracks initiate earlier compared to forward cracks observed in the 2:1 and 8:5 variants. However, the extreme stretchability afforded by the sideways fracture mechanism prevents full rupture, allowing the material to accommodate large deformations without complete failure.

\begin{table}[H]
    \centering
    \caption{\textbf{Experimental results for the displacement and average engineering strain at the onset of crack extension.} The displacement at crack onset corresponds to the instant when initial crack growth take place---and not when full rupture occurs, which is reported in Figure \ref{fig:barplots_exp}. The results are identified from the images for the four experimental repetitions (denoted as ``Rep.''). The mean values and standard deviation for each material are included. The average engineering strain at crack onset is obtained dividing the value of displacement of the clamps by the initial distance between clamps ($l_\text{0}$).}
    \begin{tabular}{l | c | c | c | c | c | c | c}
         & Rep. & \cellcolor[HTML]{F5F5F5} Elastosil 2:1 & \cellcolor[HTML]{F5F5F5} Elastosil 8:5 & \cellcolor[HTML]{F5F5F5} Elastosil 1:1 & \cellcolor[HTML]{F5F5F5} VHB & \cellcolor[HTML]{F5F5F5} Sylgard \\
        \midrule
        \multirow{5}{*}{\begin{tabular}[l]{@{}c@{}} Displacement at\\crack onset [mm]\end{tabular}} & 1 & \qty{15.6}{} & \qty{36.3}{} & \qty{27.5}{} & \qty{42.5}{} & \qty{6.3}{} \\
         & 2 & \qty{36.8}{} & \qty{31.1}{} & \qty{28.3}{} & \qty{65.2}{} & \qty{5.2}{} \\
         & 3 & \qty{24.9}{} & \qty{28.3}{} & \qty{22.1}{} & \qty{57.8}{} & \qty{5.3}{} \\
         & 4 & \qty{32.6}{} & \qty{23.5}{} & \qty{26.9}{} & \qty{29.8}{} & \qty{5.7}{} \\
         \cline{2-7}
         &  & $27.5 \pm 8.08$ & $29.8 \pm 4.63$ & $26.2 \pm 2.42$ & $ 48.8 \pm 13.7$ & $5.6 \pm 0.4 $ \\
         \midrule
         \multirow{4}{*}{\begin{tabular}[l]{@{}c@{}} Average eng. strain at\\crack onset [\%]\end{tabular}} & 1 & \qty{18.4}{} & \qty{42.7}{} & \qty{32.4}{} & \qty{50.0}{} & \qty{7.4}{} \\
         & 2 & \qty{43.3}{} & \qty{36.6}{} & \qty{33.3}{} & \qty{76.7}{} & \qty{6.1}{} \\
         & 3 & \qty{29.3}{} & \qty{33.3}{} & \qty{26.0}{} & \qty{68.0}{} & \qty{6.2}{} \\
         & 4 & \qty{38.4}{} & \qty{27.6}{} & \qty{31.6}{} & \qty{35.1}{} & \qty{6.7}{} \\
         \cline{2-7}
         &  & $32.4 \pm 9.5$ & $35.1 \pm 5.5$ & $30.8 \pm 2.9$ & $57.5 \pm 16.1$ & $6.6 \pm 0.5$ \\
    \end{tabular}
    \label{tab:displacement_crack_onset}
\end{table}

\begin{figure}[H]
\centering
\includegraphics[width=0.7\textwidth]{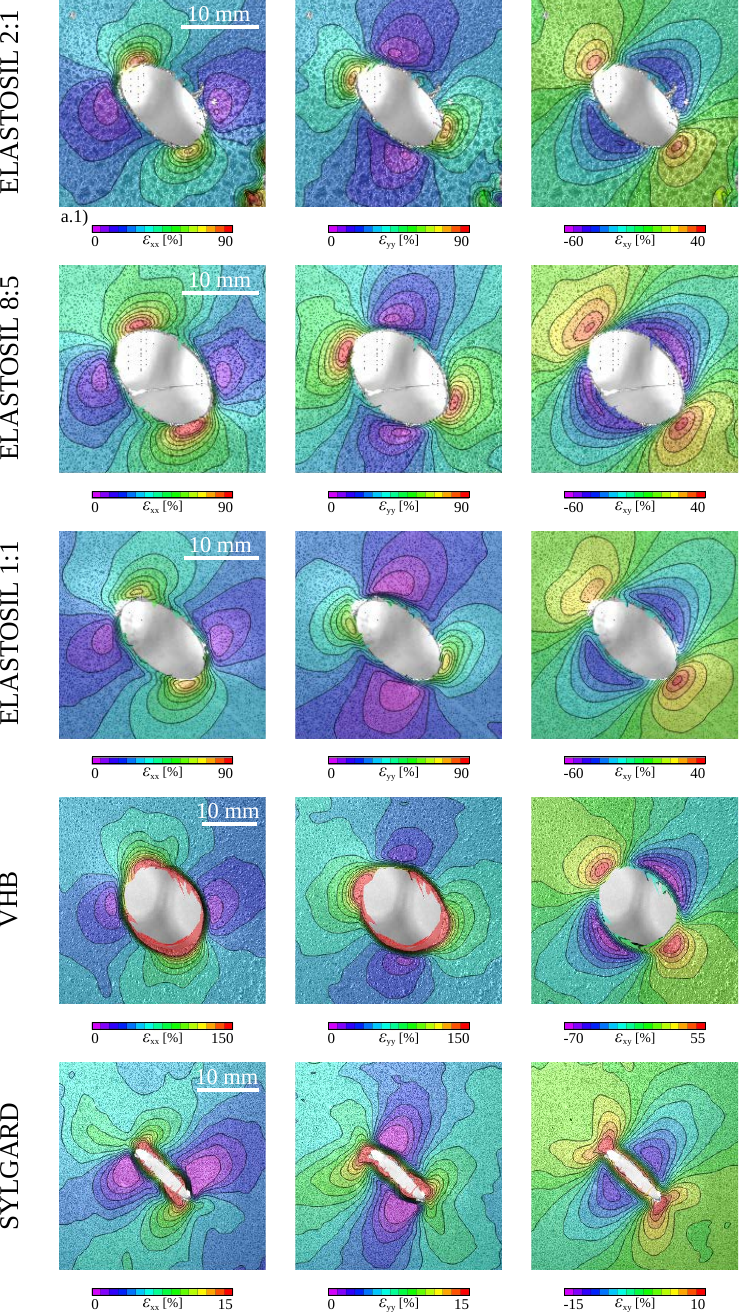}
\caption{\textbf{Strain fields at crack onset.} The fields are engineering strain fields computed from the Lagrange strain tensor. 
The fields correspond to the loading stages described in Figure~\ref{fig:Force-Disp}.a.1-5 and in Table~\ref{tab:displacement_crack_onset}. The fields correspond for the first repetition (out of the four repetitions performed for the same type of sample and same test conditions). Note that $x$ and $y$ denote the horizontal and  vertical directions, respectively.}
\label{fig:FigDICStrains}
\end{figure}

Until now, the discussion has focused on the average strain experienced by the samples. However, the local deformation near the central notch offers additional insight into the fracture behavior. To explore this, we use DIC to analyze the strain fields in the vicinity of the crack tip. Figure~\ref{fig:FigDICStrains} presents the full-field strain maps at the onset of crack extension (cf. Table~\ref{tab:displacement_crack_onset}) for experimental repetition 1 across all materials. The corresponding strain fields for repetitions 2, 3, and 4 are provided in \ref{sec:Additional_Strain_Fields} (Figures~\ref{fig:FigDICStrains2}–\ref{fig:FigDICStrains4}). Consistent with the global metrics discussed earlier (total work of fracture and average strain at failure), VHB exhibits the highest local strains at the onset of crack extension, reaching values of \qty{100}{\%} in both $\epsilon_{xx}$ and $\epsilon_{yy}$\footnote{The strain fields display symmetry with respect to the central notch orientation.}. The shear component $\epsilon_{xy}$ peaks at \qty{55}{\%}. In contrast, Sylgard shows the lowest crack-tip strains, with peak values of \qty{15}{\%} for $\epsilon_{xx}$ and $\epsilon_{yy}$, and \qty{10}{\%} for $\epsilon_{xy}$. Notably, the ratio $\epsilon_{xy} / \epsilon_{xx/yy}$ is higher for Sylgard than for VHB. This is attributed to the sharper crack geometry in Sylgard resulting from it higher stiffness, which concentrates strain more intensely compared to the blunter, more rounded cracks seen in Elastosil and VHB, where large deformations cause the crack to evolve into a circumferential-like cavity. For the Elastosil materials, the low-crosslinked version (Elastosil 2:1) reaches strain values of \qty{78}{\%} in $\epsilon_{xx}$ and $\epsilon_{yy}$, and \qty{30}{\%} in $\epsilon_{xy}$. The intermediate-crosslinked variant (Elastosil 8:5) displays slightly higher values: \qty{85}{\%}, \qty{85}{\%}, and \qty{34}{\%}, respectively. Finally, the highly crosslinked formulation (Elastosil 1:1), which undergoes sideways fracture, exhibits strains of \qty{73}{\%} ($\epsilon_{xx}$), \qty{70}{\%} ($\epsilon_{yy}$), and \qty{34}{\%} ($\epsilon_{xy}$). As discussed previously, these comparatively lower values can be explained by the earlier onset of crack propagation in Elastosil 1:1. Nonetheless, the sideways fracture mechanism permits large deformations without complete rupture. In summary, the local strain fields offer valuable insight into the distribution and concentration of deformation at the crack tip. However, they are insufficient to fully describe the material’s mechanical response, particularly in terms of stress. To address this, the next section introduces a data-adaptive framework for hyperelastic energy functions capable of capturing the stress–strain behavior underlying the observed deformation patterns for different soft constitutive behaviors.

\section{Data-adaptive hyperelastic constitutive modeling}
\begin{figure}[H]
\centering
\includegraphics[width=0.9\textwidth]{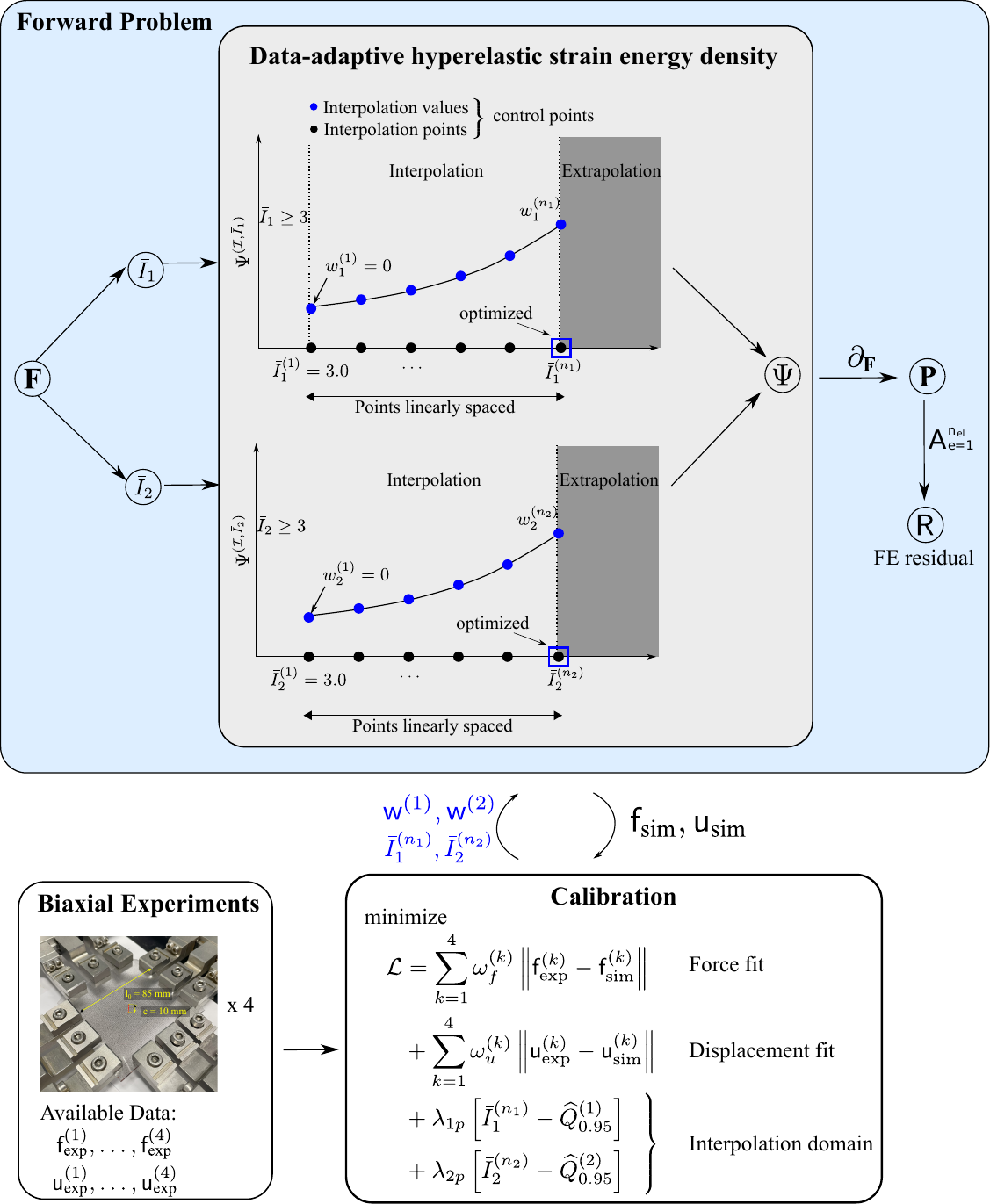}
\caption{\textbf{Schematics of the data-adaptive hyperelastic constitutive modeling framework.} The invariants $\firstInvariant, \secondInvariant$ are discretized with interpolation points (black circles) and the corresponding interpolation values (blue circles) represent unknown material parameters. Interpolation points and values uniquely define the B-Spline interpolation functions, which, if required, can be extrapolated by continuing the last spline segment. The interpolation values and the interpolation domain are determined via calibration, using global reaction forces and DIC data obtained from biaxial experiments.}
\label{fig:FigSchematics}
\end{figure}

Our framework for data-adaptive-based configurational forces for fracture is built on two key pillars: i) a data-adaptive framework for hyperelastic energy functions and ii) the Configurational Force Method, implemented as a postpocessing algorithm. The present section sets the stage building up on the data-adaptive framework developed by the authors that captures the constitutive behavior with sufficient accuracy (see seminal works \cite{Wiesheier2023,Wiesheier2024})---because accurately capturing the constitutive behavior of soft solids is key for a reliable application of the Configurational Force Method to soft solids. In this regard, some challenges were outlined by the authors in \cite{Moreno-Mateos2024b}. The framework integrates reaction force measurements from the testing machine with displacement fields obtained via DIC, and remains robust even when the DIC data is spatially incomplete, i.e., limited to a region of interest (ROI)\footnote{Our data-driven hyperelastic energy functions are obtained with a finite element updating method that provides robustness in incorporating information from displacement fields spatially limited. The material parameters in the weak form are fixed and the displacement field is solved. Then, the material parameters are updated to minimize a residual set with the experiments. This residual contains full field data (displacement field) and force-displacement (reaction forces) data.}. Figure \ref{fig:FigSchematics} provides an overview of the methodology. The hyperelastic strain energy functions will be eventually used in the calculation of the Eshelby stress tensor in the Configurational Force Method. Our FEMU optimization strategy requires the repeated solution of the mechanical boundary-value problem (forward problem), derived succinctly in the following.

\newpage

\subsection{Forward problem: kinematics, strain energy density, stress tensor, FE-implementation}
In finite deformation theory, a generic point $\positionMat \in \bodyMat$ in the material configuration $\bodyMat$ is mapped to the deformed configuration $\body$ via the deformation map $\deformationMap$, i.e., $\positionSpt = \boldsymbol \Phi \left(\positionMat\right)$. The gradient of the mapping with respect to the material coordinates is defined as the second-order deformation gradient tensor
\begin{align}
    \defGrad = \frac{\partial \positionSpt}{\partial \positionMat}, \qquad  \detDefGrad = \det \defGrad  > 0,
\end{align}
where $\detDefGrad$ denotes the determinant of the deformation gradient. The deformation is decomposed into volumetric and isochoric contributions according to
\begin{align}
    \defGrad = \volumetric{\defGrad} \isochoric{\defGrad}, \quad \volumetric{\defGrad} = \detDefGrad^{-1/3}\secondOrderUnitTensorMat , \quad \isochoric{\defGrad} = \detDefGrad^{1/3} \defGrad,
\end{align}
with $\secondOrderUnitTensorMat$ denoting the second-order unit tensor. This multiplicative decomposition carries over to the right Cauchy-Green tensor
\begin{align}
    \rightCauchyGreen = \transpose{\defGrad} \defGrad, \quad \isochoric{\rightCauchyGreen} = \detDefGrad^{-2/3} \rightCauchyGreen.
\end{align}
Per construction, $\det(\isochoric{\rightCauchyGreen}) = 1$, hence, there are only two principal invariants $\firstInvariant = \trace{\isochoric{\rightCauchyGreen}}, \; \secondInvariant = \trace{\cofactor{\isochoric{\rightCauchyGreen}}}$ characterizing the isotropic hyperelastic material response. These invariants serve as arguments of the strain energy density (per unit volume) given by
\begin{align}
    \strainEnergyFunction = \pressure[\detDefGrad-1] + \isochoric{\strainEnergyFunction}(\firstInvariant, \secondInvariant). 
\end{align}
In this expression, $\pressure$ is a Lagrange multiplier. In particular, $\pressure$ is not derived from a volumetric strain energy density, as commonly approached when modeling compressible material behavior. The material must be thermodynamically consistent. This is automatically satisfied when the Piola stress tensor is derived from the strain energy density:
\begin{align}
    \pkStress = \frac{\partial \isochoric{\strainEnergyFunction}}{\partial \defGrad} + \pressure \invTrans{\defGrad}.
\end{align}  
Neglecting body forces, the mechanical boundary-value problem has the format:
\begin{equation}
\begin{alignedat}{2}
\Div{\pkStress} &= \mathbf{0} \quad &&\text{in } \bodyMat, \\
\det{\defGrad} - 1 &= 0 \quad &&\text{in } \bodyMat, \\
\displacement &= \widetilde{\displacement} \quad &&\text{on } \bodyBoundaryMat^{\mathrm{D}}, \\
\pkStress \mathbf{N} &= \widetilde{\mathbf{T}} \quad &&\text{on } \bodyBoundaryMat^{\mathrm{N}}.
\end{alignedat}
\label{eq:EqStrongForm}
\end{equation}
This formulation suggests there are two equations to be satisfied: $\Div{\pkStress} = \mathbf{0}$ represents the balance of linear momentum in the material configuration, and $\detDefGrad - 1 = 0$ accounts for incompressibility of the considered materials. Together, these lead to a mixed FE-formulation, where the pressure $\pressure$ and the displacement $\displacement$ are the primary nodal unknowns. We discretize $\displacement$ using linear Lagrange polynomials, and $\pressure$ is element-wise constant. As a result, the FE residual vector splits into two blocks: $\residualVectorFE{} = \left[\residualVectorFE{\displacement}, \residualVectorFE{\pressure} \right]$. The notation of the FE residual vector is important for the formulation of the optimization problem for calibrating our data-adaptive hyperelastic energy functions, as will be shown later.

\subsection{B-Spline interpolation strain energy density}
As the pressure field $\pressure$ does not originate from a volumetric strain energy density, material parameters are exclusively confined to the isochoric strain energy density. Instead of specifying a canonical model (e.g, Mooney-Rivlin \cite{mooney_theory_1940} or Yeoh \cite{yeoh_characterization_1990}), we express the isochoric strain energy density using an additive B-Spline interpolation representation,
\begin{align}
    \interpolantVarIso{\strainEnergyFunction}{\firstInvariant, \secondInvariant} = 
    \underbrace{\sum\limits_{i=1}^{\firstInvariantPts} \controlPoint{i}{1} \bspline{i}{1}(\firstInvariant)}_{\interpolantVarIso{\strainEnergyFunction}{\firstInvariant}} + 
    \underbrace{\sum\limits_{j=1}^{\secondInvariantPts} \controlPoint{i}{2} \bspline{j}{2}(\secondInvariant)}_{{\interpolantVarIso{\strainEnergyFunction}{\secondInvariant}}},
    \label{eq:Ansatz}
\end{align}
where $\interpolantVarIso{\strainEnergyFunction}{\firstInvariant}, \interpolantVarIso{\strainEnergyFunction}{\secondInvariant}$ are univariate cubic B-Spline function with control points
\begin{align}
    \controlPointVec{1} = 
    \left[\controlPoint{1}{1},\dots,\controlPoint{\firstInvariantPts}{1}\right], \quad
    \controlPointVec{2} = \left[\controlPoint{1}{2},\dots,\controlPoint{\secondInvariantPts}{2}\right].
\end{align}
Cubic B-Spline basis functions were chosen as the integration into an FE code requires at least continuous second derivatives of the univariate splines. The splines are constructed from prescribed interpolation points and corresponding interpolation values, see Figure~\ref{fig:FigSchematics} for a graphical representation on the example of the first invariant $\firstInvariant$. Specifically, we define
\begin{align}
    &\text{Interpolation points:} 
    && \left\{ \firstInvariantPoint{i} \right\}_{i=1}^{\firstInvariantPts}, \quad 
       \left\{ \secondInvariantPoint{i} \right\}_{i=1}^{\secondInvariantPts}. \notag \\
    &\text{Interpolation values:} 
    && \interpPointVec{1} = 
       \left[ \interpolantVarIso{\strainEnergyFunction}{\firstInvariant}(\firstInvariantPoint{i}) \right]_{i=1}^{\firstInvariantPts} 
       = \left[ \interpPoint{1}{1}, \dots, \interpPoint{1}{\firstInvariantPts} \right], \notag \\
    &&& \interpPointVec{2} = 
       \left[ \interpolantVarIso{\strainEnergyFunction}{\secondInvariant}(\secondInvariantPoint{i}) \right]_{i=1}^{\secondInvariantPts} 
       = \left[ \interpPoint{2}{1}, \dots, \interpPoint{2}{\secondInvariantPts} \right] \notag.
\end{align}

This implies that the splines are uniquely defined in terms of the interpolation points and values. The control points, as appearing in Equation~\eqref{eq:Ansatz}, are computed by solving a linear equation system in a pre-processing stage in the forward solve. Computational costs associated with the control point computation are negligible. We refer to \cite{d_boor_practical_1978, piegl_nurbs_1996} for more insights into splines.

Several requirements must be met to use the splines for describing a hyperelastic material response:
\begin{itemize}
    \item \textbf{Thermodynamic consistency:} 
    $\pkStress = \dfrac{\partial \interpolantVarIso{\strainEnergyFunction}{\firstInvariant, \secondInvariant}}{\partial \defGrad} $.
    \item \textbf{Material frame indifference:}
    $\interpolantVarIso{\strainEnergyFunction}{\firstInvariant, \secondInvariant}(\defGrad \mathbf{Q}) = \interpolantVarIso{\strainEnergyFunction}{\firstInvariant, \secondInvariant}(\defGrad) \quad \forall \mathbf{Q} \in \mathrm{SO}(3)$.
    \item \textbf{Material symmetry:}
    $\interpolantVarIso{\strainEnergyFunction}{\firstInvariant, \secondInvariant}(\mathbf{Q} \defGrad) = \interpolantVarIso{\strainEnergyFunction}{\firstInvariant, \secondInvariant}(\defGrad) \quad \forall \mathbf{Q} \in \mathrm{SO}(3)$.
    \item \textbf{Positivity:}
    $\interpolantVarIso{\strainEnergyFunction}{\firstInvariant, \secondInvariant} \geq 0 \quad \forall \defGrad$.
    \item \textbf{Zero energy undeformed state:}
    $\interpolantVarIso{\strainEnergyFunction}{\firstInvariant, \secondInvariant}(\defGrad = \secUnitTensor) = 0$.
\end{itemize}
Thermodynamic consistency, material frame indifference (objectivity), and material symmetry are satisfied a priori by (i) deriving the Piola stress from a strain energy density and (ii) parameterizing it in terms of the principal invariants $\firstInvariant, \secondInvariant$. Zero energy in the undeformed state can be easily fulfilled by placing the first interpolation points at ($\defGrad = \secUnitTensor \rightarrow \firstInvariant = \secondInvariant = 3.0$) and fixing the corresponding two interpolation values to zero:
\begin{align}
    \firstInvariantPoint{1} = \secondInvariantPoint{1} = 3.0, \quad \interpPoint{1}{1} = \interpPoint{2}{1} = 0.
\end{align}
During calibration, this represents a simple bound constraint guaranteed to be satisfied at every iteration. Note that $\interpolantVarIso{\strainEnergyFunction}{\firstInvariant, \secondInvariant}(\defGrad = \secUnitTensor) = 0$ implies $\pkStress(\defGrad = \secUnitTensor) = \mathbf{0}$ due to the choice of invariants, c.f. \cite{Dammass2025}. 

\begin{remark}
There is one subtle detail concerning the spline interpolation. While the evaluations of the splines at points inside the interpolation domain are straightforward, the evaluations at points outside the interpolation domain, i.e., out-of-bounds evaluations, need to be adequately addressed. To that end, we implemented a robust extrapolation strategy by continuing the last spline segment. This allows for evaluation at arbitrary points $\firstInvariant, \secondInvariant$ in the invariant space, regardless of the interpolation domain. Nevertheless, we want to emphasize that extrapolation should not occur frequently. Hence, it is desirable to tie the interpolation domain to the sampled invariants. This is guaranteed during the calibration stage of our data-adaptive hyperelastic energy functions, discussed in the following.
\end{remark}

\subsection{Material model calibration (Inverse problem)}
\begin{figure}[H]
\centering
\includegraphics[width=0.7\textwidth]{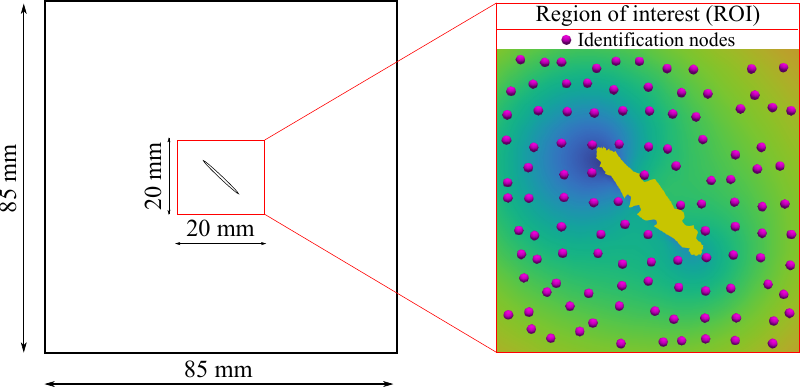}
\caption{\textbf{Region of interest used to sample the experimental displacement fields to the FE mesh and calibrate data-adaptive strain energy functions.} A square region of interest (ROI) with dimensions \qty{20}{\milli\meter} x \qty{20}{\milli\meter} in the vicinity of the crack contour is defined. The ROI is chosen consistently for each material, except for Elastosil 8:5, where the ROI has dimensions \qty{17}{\milli\meter} x \qty{17}{\milli\meter}. For the material model calibration, the DIC information is restricted to the highlighted identification nodes within the ROI. These were obtained by defining $10\times10$ evenly distributed sampling points in the ROI and finding the closest node of the FE-mesh for each sampling point. The resulting FE-mesh nodes define the set of identification nodes. The yellow-colored area in the center of the ROI indicates missing DIC data -- this area can be different for each repetition.}
\label{fig:FigROI}
\end{figure}

The unknown interpolation values are determined via non-linear least squares optimization. This procedure requires the repeated solution of the bespoke forward problem until a stopping criterion is met. 
Our data-adaptive framework leverages full-field displacement fields obtained from DIC and global reaction forces across four experimental repetitions for each material. This compreshensive experimental database motivates the following objective function:
\begin{align}
    \objectiveFunc = &\sum\limits_{k=1}^{4} 
    \left[ 
    \forceWeight{k} \| \forceVec{\text{sim}}{k} - \forceVec{\text{exp}}{k} \|^2 + 
    \dispWeight{k} \| \dispVec{\text{sim}}{k} - \dispVec{\text{exp}}{k} \|^2
    \right] + \\
    &\quantileWeight{1p} \left[ \firstInvariantPoint{\firstInvariantPts} - \quantile{0.95}{1} \right] + 
    \quantileWeight{2p} \left[ \secondInvariantPoint{\secondInvariantPts} - \quantile{0.95}{2} \right].
    \label{eq:EqLoss}
\end{align}
This formulation suggests that all available data per material are aggregated such that there is one strain energy density for each material. The inclusion of physical quantities with different units (forces and displacements) necessitates an appropriate weighting strategy. Specifically, the individual contributions of the objective function are weighted according to
\begin{align}
    \forceWeight{k} = \frac{1}{\max(\forceVec{\text{exp}}{k})}, \quad \dispWeight{k} = \frac{1}{\max(\dispVec{\text{exp}}{k})},
\end{align}
to account for the different scales of forces and displacements, and normalization of the data. The vector $\forceVec{\text{exp}}{k}$ contains measured reaction forces at multiple loading stages of the clamps, and the simulation counterpart $\forceVec{\text{sim}}{k}$ represents the numerical reaction force at the equivalent load step in the forward model. Similarly, $\dispVec{\text{exp}}{k}$ assembles measured displacements at multiple clamp loading stages. Bilinear interpolation in space using the \texttt{ResampleWithDataset} (available in Paraview) is performed to interpolate the measured displacements onto the nodes of the FE-mesh. Additionally, the measured displacements are downsampled to identification nodes defined within a ROI in the crack vicinity (cf. Figure \ref{fig:FigROI} for a qualitative illustration). The identification nodes were chosen consistently across the four experimental repetitions; that is, the spatial positions of incorporated displacements are identical for each repetition.

Aside from the choice of experimental data for the calibration, another important detail of the calibration is the definition of the interpolation domain for the B-Spline interpolation functions. During the forward problem, the splines are evaluated at the invariant values at the quadrature points. However, these evaluation points are unknown before the calibration and are implicitly defined by the deformation gradient. This impedes the definition of fixed interpolation points for the calibration and suggests considering the interpolation domains also as unknown. That said, we expand the parameter space by treating the last two interpolation points $\firstInvariantPoint{\firstInvariantPts}, \secondInvariantPoint{\secondInvariantPts}$ as additional optimization variables, along with the interpolation values. This ensures that the interpolation domain aligns well with the sampled invariants, but necessitates the inclusion of two additional loss terms to drive $\firstInvariantPoint{\firstInvariantPts}, \secondInvariantPoint{\secondInvariantPts}$ towards the sampled invariant space, approximated by the 95th-percentiles $\quantile{0.95}{j}$, obtained from Kernel Density estimation (KDE). More details on KDE are presented in \ref{sec:SecKernel}. 

Eventually, the parameter optimization problem has the format
\begin{align}
\theta &=
\left[
\interpPointVec{1},\;
\interpPointVec{2},\;
\firstInvariantPoint{\firstInvariantPts},\;
\secondInvariantPoint{\secondInvariantPts}
\right], \notag \\[0.5em]
\displaystyle \theta^\ast &=
\mathop{\arg \min}_{\theta \in \mathbb{R}^{\firstInvariantPts + \secondInvariantPts + 2}} \;
\objectiveFunc, \notag \\[0.5em]
\text{subject to} \quad
&\frac{\mathrm{d}^2 \interpolantVarIso{\strainEnergyFunction}{\firstInvariant}}{\mathrm{d} \firstInvariant^2} \geq 0, \notag \\ 
&\frac{\mathrm{d}^2 \interpolantVarIso{\strainEnergyFunction}{\secondInvariant}}{\mathrm{d} \secondInvariant^2} \geq 0, \notag \\
&\residualVectorFE{}(\theta) = \mathsf{0}.
\label{eq:OptimizationProblem}
\end{align}

The positivity of the second derivatives with respect to the invariants constitutes a set of nonlinear inequality constraints, handled by the optimization solver with Lagrange multipliers. Conversely, $\residualVectorFE{}(\theta) = 0$ ensures that the equilibrium conditions (weak form of the balance of linear momentum, incompressibility) are satisfied. We solve this constrained optimization problem using \textit{fmincon}, the nonlinear programming solver provided by \textit{MATLAB}. At every iteration, the objective function is evaluated by calling our in-house forward solver based on the \texttt{deal.II} library \cite{arndt_dealii_nodate, arndt_dealii_2021}. The optimization terminates when the relative change in either the objective function value or the parameter step size between successive iterations falls below the tolerance \qty{1e-12}{}.

\newpage

\subsection{Discovered material models}
\begin{figure}[H]
\centering
\includegraphics[width=0.85\textwidth]{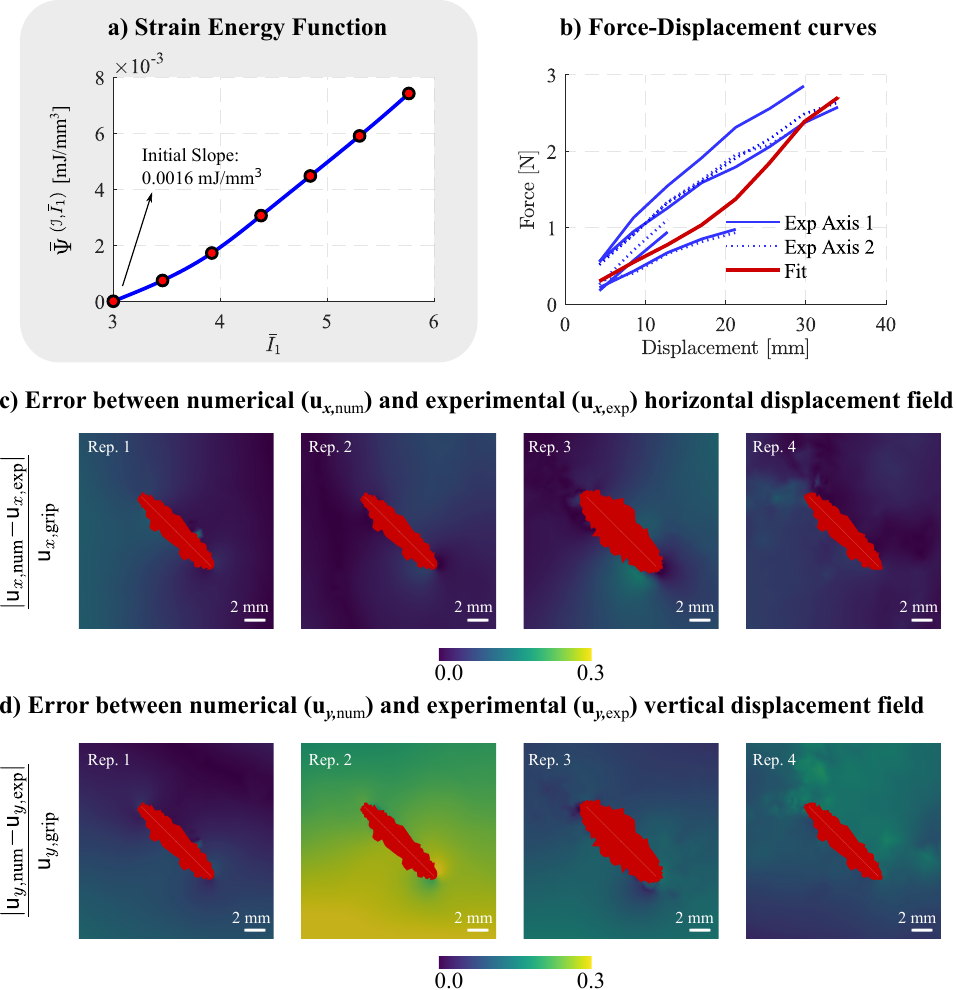}
\caption{\textbf{Data-adaptive strain energy function for Elastosil 2:1 obtained from biaxial experimental data: reaction forces and full-field DIC measurements.} 
(a) Data-adaptive hyperelastic strain energy density as a function of the first invariant, $\firstInvariant$. The energy function is reconstructed as a linear combination of control points defined in the invariant space and B-spline basis functions (see Equation~\ref{eq:Ansatz}), with a maximum value of $\firstInvariant$ of \qty{5.76}{}.
(b) Comparison between experimental and numerically fitted reaction forces at the clamps, plotted against the displacement of the biaxial testing machine’s actuators (total displacement per axis).
(c) Normalized error in the horizontal displacement field within the region of interest—specifically the crack and crack-tip vicinity (cf. Figure~\ref{fig:FigROI})—across four experimental repetitions under identical loading conditions. Displacement of the grips is used as the normalization reference.
(d) Normalized error in the vertical displacement field for the same region of interest and experimental repetitions as in (c), using the same normalization convention.
}
\label{fig:FigElastosil21}
\end{figure}

\begin{figure}[H]
\centering
\includegraphics[width=0.85\textwidth]{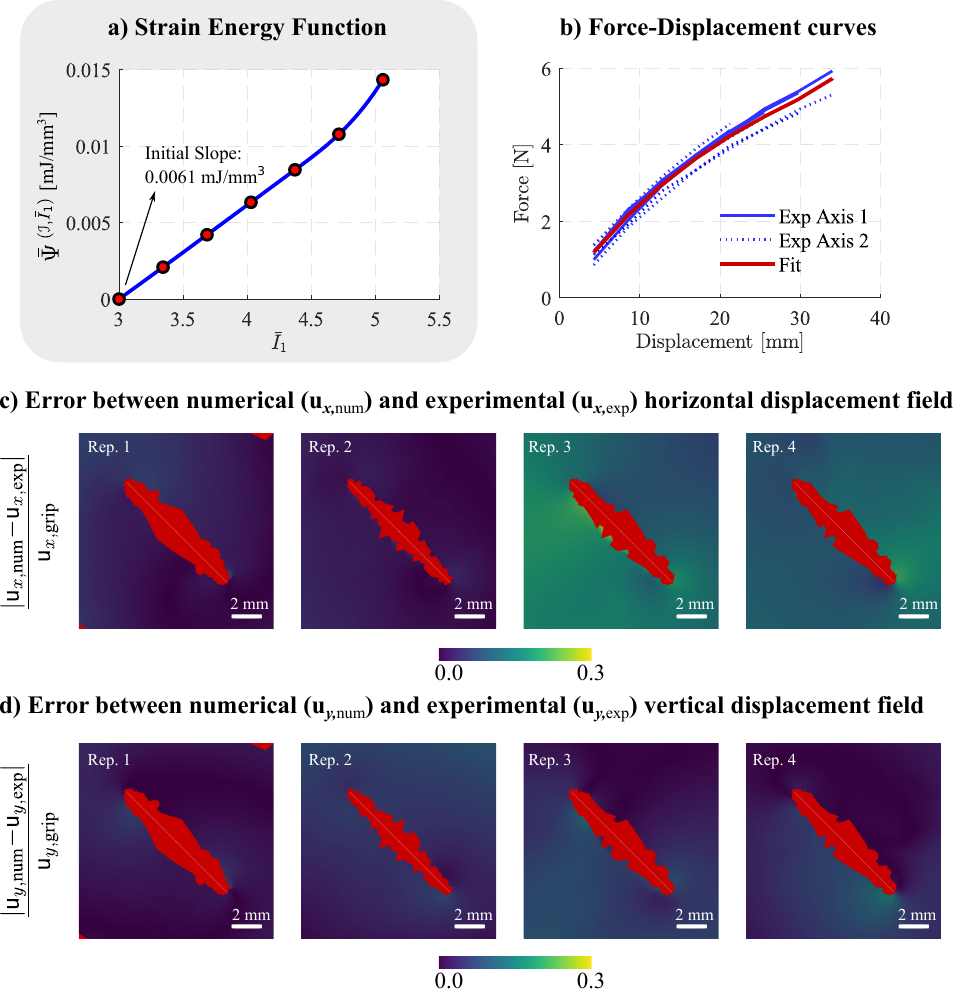}
\caption{\textbf{Data-adaptive strain energy function for Elastosil 8:5 obtained from biaxial experimental data: reaction forces and full-field DIC measurements.} 
(a) Data-adaptive hyperelastic strain energy density as a function of the first invariant, $\firstInvariant$. The energy function is reconstructed as a linear combination of control points defined in the invariant space and B-spline basis functions (see Equation~\ref{eq:Ansatz}), with a maximum value of $\firstInvariant$ of \qty{5.06}{}.
(b) Comparison between experimental and numerically fitted reaction forces at the clamps, plotted against the displacement of the biaxial testing machine’s actuators (total displacement per axis).
(c) Normalized error in the horizontal displacement field within the region of interest—specifically the crack and crack-tip vicinity (cf. Figure~\ref{fig:FigROI})—across four experimental repetitions under identical loading conditions. Displacement of the grips is used as the normalization reference.
(d) Normalized error in the vertical displacement field for the same region of interest and experimental repetitions as in (c), using the same normalization convention.
}
\label{fig:FigElastosil85}
\end{figure}

\begin{figure}[H]
\centering
\includegraphics[width=0.85\textwidth]{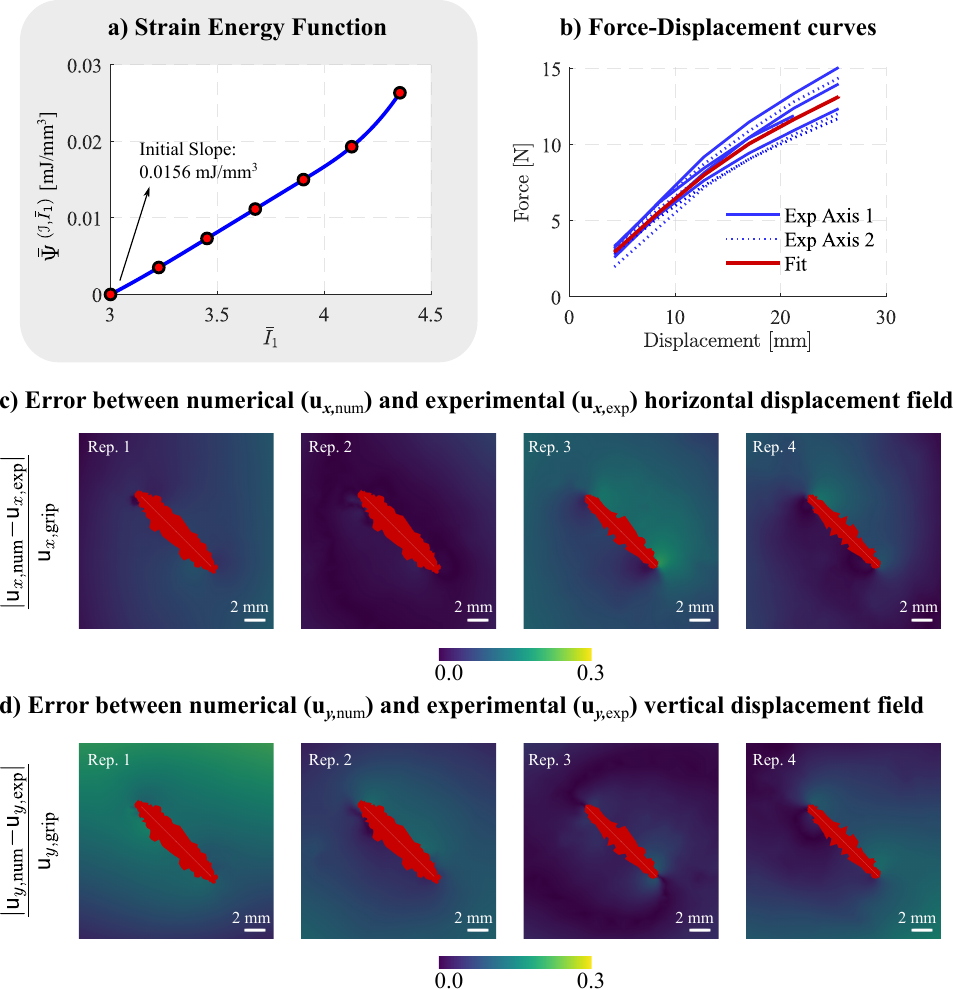}
\caption{\textbf{Data-adaptive strain energy function for Elastosil 1:1 obtained from biaxial experimental data: reaction forces and full-field DIC measurements.} 
(a) Data-adaptive hyperelastic strain energy density as a function of the first invariant, $\firstInvariant$. The energy function is reconstructed as a linear combination of control points defined in the invariant space and B-spline basis functions (see Equation~\ref{eq:Ansatz}), with a maximum value of $\firstInvariant$ of \qty{4.35}{}.
(b) Comparison between experimental and numerically fitted reaction forces at the clamps, plotted against the displacement of the biaxial testing machine’s actuators (total displacement per axis).
(c) Normalized error in the horizontal displacement field within the region of interest—specifically the crack and crack-tip vicinity (cf. Figure~\ref{fig:FigROI})—across four experimental repetitions under identical loading conditions. Displacement of the grips is used as the normalization reference.
(d) Normalized error in the vertical displacement field for the same region of interest and experimental repetitions as in (c), using the same normalization convention.}
\label{fig:FigElastosil11}
\end{figure}

\begin{figure}[H]
\centering
\includegraphics[width=0.85\textwidth]{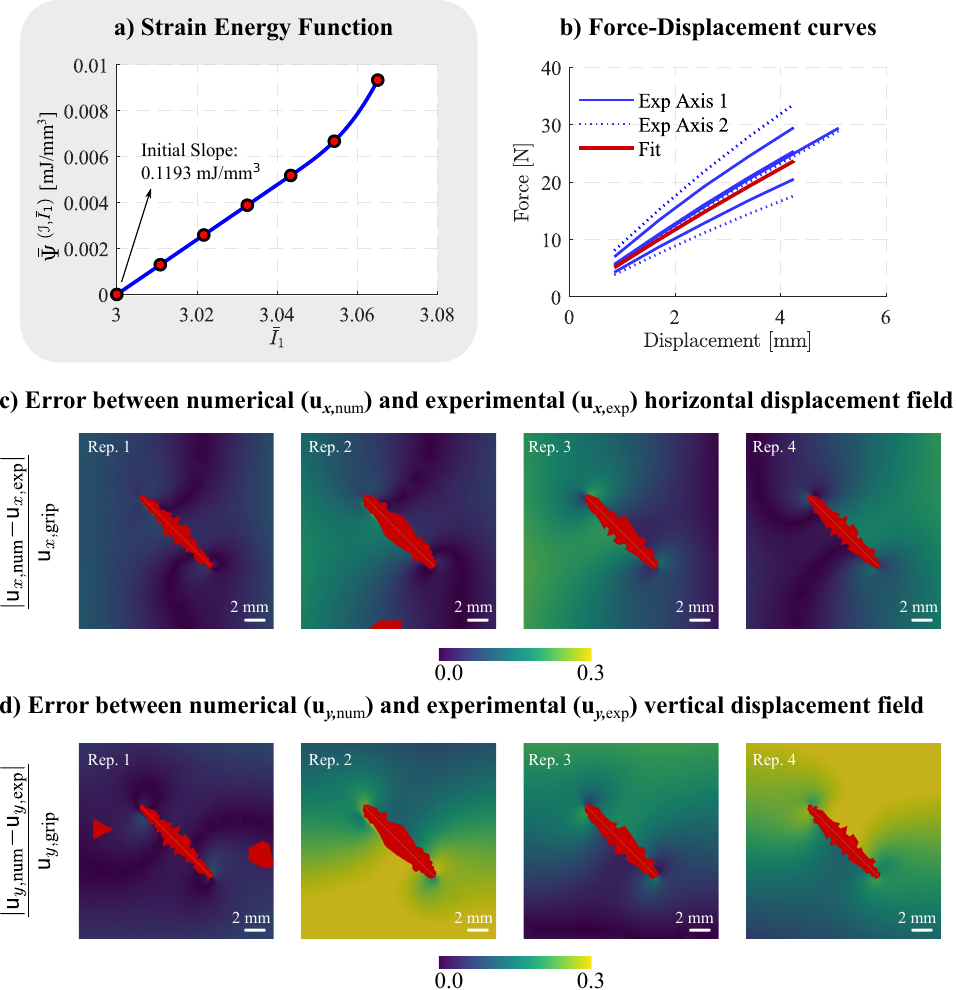}
\caption{\textbf{Data-adaptive strain energy function for Sylgard obtained from biaxial experimental data: reaction forces and full-field DIC measurements.} 
(a) Data-adaptive hyperelastic strain energy density as a function of the first invariant, $\firstInvariant$. The energy function is reconstructed as a linear combination of control points defined in the invariant space and B-spline basis functions (see Equation~\ref{eq:Ansatz}), with a maximum value of $\firstInvariant$ of \qty{3.07}{}.
(b) Comparison between experimental and numerically fitted reaction forces at the clamps, plotted against the displacement of the biaxial testing machine’s actuators (total displacement per axis).
(c) Normalized error in the horizontal displacement field within the region of interest—specifically the crack and crack-tip vicinity (cf. Figure~\ref{fig:FigROI})—across four experimental repetitions under identical loading conditions. Displacement of the grips is used as the normalization reference.
(d) Normalized error in the vertical displacement field for the same region of interest and experimental repetitions as in (c), using the same normalization convention.
}
\label{fig:FigSylgard}
\end{figure}

\begin{figure}[H]
\centering
\includegraphics[width=0.85\textwidth]{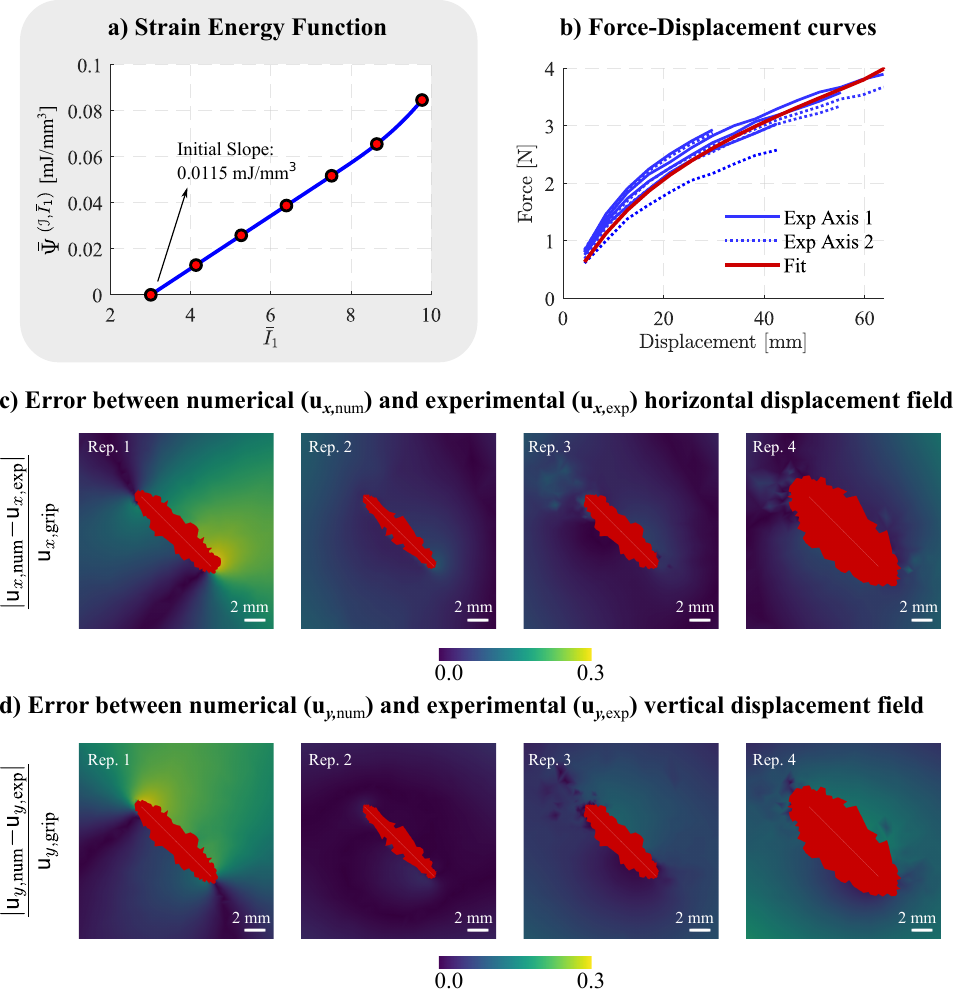}
\caption{\textbf{Data-adaptive strain energy function for VHB obtained from biaxial experimental data: reaction forces and full-field DIC measurements.} 
(a) Data-adaptive hyperelastic strain energy density as a function of the first invariant, $\firstInvariant$. The energy function is reconstructed as a linear combination of control points defined in the invariant space and B-spline basis functions (see Equation~\ref{eq:Ansatz}), with a maximum value of $\firstInvariant$ of \qty{9.77}{}.
(b) Comparison between experimental and numerically fitted reaction forces at the clamps, plotted against the displacement of the biaxial testing machine’s actuators (total displacement per axis).
(c) Normalized error in the horizontal displacement field within the region of interest—specifically the crack and crack-tip vicinity (cf. Figure~\ref{fig:FigROI})—across four experimental repetitions under identical loading conditions. Displacement of the grips is used as the normalization reference.
(d) Normalized error in the vertical displacement field for the same region of interest and experimental repetitions as in (c), using the same normalization convention.
}
\label{fig:FigVHB}
\end{figure}

Figures~\ref{fig:FigElastosil21}, \ref{fig:FigElastosil85}, \ref{fig:FigElastosil11}, \ref{fig:FigSylgard} and \ref{fig:FigVHB} summarize the results for the materials Elastosil 2:1, Elastosil 8:5, Elastosil 1:1, Sylgard, and VHB, respectively. Each figure follows a consistent layout and includes: (a) the identified strain energy density, (b) a comparison between the numerical and experimental reaction forces, and the error between numerical and experimental displacement fields in both (c) horizontal and (d) vertical directions. 

Our data-adaptive framework for hyperelastic energy functions leverages both reaction forces data and full-field displacement measurements from DIC to optimize the energy functions. To ensure accuracy, the discrepancies between experimental and computational data—both in terms of reaction forces and displacement fields—must be inspected. We leverage the identified strain energy densities to solve the forward problem and obtain the numerical reaction force (see Figures~\ref{fig:FigElastosil21}-\ref{fig:FigVHB}.b). While the experimental data comprises eight curves---four repetitions and two axes---, there is only a single numerical prediction (solid red line). The numerical reaction force consistently falls within the scatter of the experimental data, indicating good average agreement. For Elastosil 2:1, it is noteworthy that the predicted reaction force shown in Figure~\ref{fig:FigElastosil21} reflects the scatter observed in the experimental force-displacement curves, particularly in terms of the maximum displacement. As a result, at large displacements, the numerical curve tends to follow the subset of experimental curves that extend to similarly large displacement values. In turn, the bottom panels of the figures show the error between the numerical and experimental displacement field in both horizontal and vertical direction within the ROI defined in Figure~\ref{fig:FigROI}. To ensure comparability between materials, the absolute displacement error is normalized by the applied grip displacement. The red ellipsoidal regions near the center of the contour plots correspond to areas where DIC data is unavailable; these regions vary across repetitions. While some contours display larger errors, the error remains close to zero for most of the contours, including near the crack tip.

The data-adaptive optimization of the hyperelastic energy functions based on biaxial test data primarily activates the first invariant strain energy function, $\interpolantVarIso{\strainEnergyFunction}{\firstInvariant}$. As outlined in the previous section, we assume an isochoric strain energy density that depends on the invariants $\firstInvariant$ and $\secondInvariant$. During optimization, the optimizer consistently deactivated the $\secondInvariant$ contribution by setting all interpolation values ($\interpPoint{2}{j}$) to their prescribed lower bound of zero \footnote{In mathematical terms, $\interpPoint{2}{j} = 0$ indicates that the Lagrange multipliers associated with the lower bound constraints are active at the solution. This means the optimizer would have preferred a solution with $\interpPoint{2}{j} < 0$, but such values are not allowed because the strain energy contribution associated with $\secondInvariant$ must remain non-negative.}. This results in a reduced model depending solely on $\firstInvariant$. We refer to \ref{sec:SecInvariants} for an extended investigation on the invariants. Across all materials, the condition of zero energy in the undeformed state is satisfied, i.e, $\interpolantVarIso{\strainEnergyFunction}{\firstInvariant}(\firstInvariant = 3.0)$, and the interpolated curves are convex in the interpolation domain. Dammaß and co-authors \cite{Dammass2025} recently concluded that ``models based solely on $\firstInvariant$ can make sound predictions for multiaxial loadings even if parameterised only from uniaxial data'' and that ``$\secondInvariant$-only models are completely incapable in even qualitatively capturing experimental stress data at large deformations''. In line with these findings, the optimizer in our framework for data-driven hyperelastic energy is parsimonious, identifying $\firstInvariant$-only-based energy functions as sufficient for accurate modeling. We acknowledge, however, that this conclusion is \textit{ad hoc} for our distinct data-driven optimization methodology and experimental dataset on samples with the initial central pre-cut as strain concentrator. The constitutive behaviour of our experimental data is sufficiently well described by $\firstInvariant$ and it might be that $\secondInvariant$ would slightly enhance the model performance for larger stretches.

The different fracture toughness and constitutive behavior of the elastomers yield different deformation states for crack onsets, which limits the interpolation domain for $\firstInvariant$. The interpolation domain varies across materials, as seen in Figures~\ref{fig:FigElastosil21}, \ref{fig:FigElastosil85}, \ref{fig:FigElastosil11}, \ref{fig:FigSylgard}, and \ref{fig:FigVHB}. Sylgard exhibits the smallest interpolation domain, with a maximum value of $\firstInvariant$ of \qty{3.07}{}, while VHB has the largest value, i.e., \qty{9.77}{}. For the three versions of Elastosil---2:1, 8:5, and 1:1--- the values are \qty{5.76}{}, \qty{5.06}{}, and \qty{4.35}{}, respectively. This is consistent with the experimental findings, namely the average strains at crack onset, i.e.,
\qty{6.6}{\%} for Sylgard, \qty{57.5}{\%} for VHB, \qty{32.4}{\%} for Elastosil 2:1, and \qty{35.1}{\%} for Elastosil 8:5. As discussed, no average strain at crack onset is available for Elastosil 1:1. This demonstrates the adaptability of our model in aligning the interpolation domain with the loading regimes encountered in the experiments. 

The data-adaptive hyperelastic energy functions adjust both their values and curvature to accurately reflect the distinct constitutive behaviors of the five soft elastomers studied. The identified functions $\interpolantVarIso{\strainEnergyFunction}{\firstInvariant}$ exhibit an initial linear increase, followed by a gradual increase in curvature toward the upper end of the domain. However, the slope and the curvature evolution differ significantly across the materials. Sylgard, being the most brittle material, exhibits the highest initial slope value (\qty{0.1193}{\newton \per \milli \meter \squared}), followed by VHB (\qty{0.0115}{\newton \per \milli \meter \squared}) and Elastosil 1:1 (\qty{0.0156}{\newton \per \milli \meter \squared}). The lowest values are observed for Elastosil 2:1 and Elastosil 8:5, with \qty{0.0016}{\newton \per \milli \meter \squared} and \qty{0.0061}{\newton \per \milli \meter \squared}, respectively. Although these values cannot be directly interpreted as conventional material parameters such as Young’s modulus, they can be related to the shear modulus under the assumption of a neo-Hookean material model. In this context, the (initial) slope of the strain energy function provides an estimate of the shear modulus, with the slope being half the shear modulus\footnote{Readers interested in the shear modulus for Elastosil in its different crosslinked variants may benchmark the results presented in this paper against the experimental uniaxial tensile data reported in the authors' previous work \cite{Moreno-Mateos2024n}, albeit the reported metric there was the Young’s modulus rather than the shear modulus. The results are consistent and further validate the initial slopes of the strain energy functions in this work.}. Further, the hyperelastic strain energies reveal the larger strain hardening for Elastosil 8:5 and 1:1, as well as for Sylgard, with an increase in the curvature on the upper limit of the $\firstInvariant$ interpolation domain. Overall, our data-adaptive framework for hyperelastic energy functions demonstrates versatility in capturing biaxial experimental responses across soft materials with varying stiffness and distinct hyperelastic behaviors. This is paramount because accurately capturing the constitutive behavior of soft elastomers is key for a reliable application of the Configurational Force Method to soft solids. 

\section{Configurational Force Method describes fracture onset under biaxial loading}
The framework for data-adaptive configurational forces in fracture mechanics is built upon two core components: (i) a data-adaptive formulation of hyperelastic energy functions, and (ii) the Configurational Force Method, implemented as a post-processing algorithm. The following section addresses ii) with a postprocessing algorithm that calculates the Eshelby stress tensor and its corresponding nodal forces, i.e., configurational forces. We set a computational testbed to compute configurational forces in virtual experiments that mimic the actual biaxial tests. The crack tip configurational force will be a computationally efficient estimate of the contour $J$-integral.

\subsection{The Configurational Force Method}
The kinematic description of a continuum involves establishing a relationship between the positions of physical points before and after the application of load. The parametrization of the kinematic and energetic quantities in terms of the material positions is commonly known as \textit{spatial motion problem} or rather as \textit{deformational mechanics}. In the \textit{material motion problem} or rather \textit{configurational mechanics}, the kinematic and constitutive measures are parametrized in terms of the spatial positions of the continuum via the inverse deformation gradient.

Just as spatial forces are associated with the deformation of the continuum, configurational forces drive changes in the material configuration, leading to an associated release of energy. The definition of a stress-energy tensor in the material motion description is in the heart of the theory. Akin to the energy density $\interpolantVar{\strainEnergyFunction}{\firstInvariant, \secondInvariant}\left(\mathbf{F};\mathbf{X}\right)$ defined in $\bodyMat$, an energy density $\psi=\psi\left(\mathbf{f}, \boldsymbol \phi \right)$, with $\mathbf{f}$ the inverse deformation gradient and $\phi$ the mapping function $\mathbf{X}=\boldsymbol \phi \left(\mathbf{x}\right)$ from spatial to material coordinates, can be re-defined per unit volume in $\body$, so that $\Psi \simeq \left[\det \mathbf{F}\right] \, \psi$. This new energy density serves as a potential for the Eshelby stress tensor \cite{Eshelby1951,Kienzler1997} in the material configuration through a push-back operation or, alternatively, in the form of the Eshelby energy-momentum tensor \cite{Eshelby1975} as a function of the direct deformation gradient \cite{Eshelby1999},
\begin{equation}\label{eq:Eshelby_stress_F}
\boldsymbol \Sigma\left(\mathbf{F}\right)
=\left[\interpolantVarIso{\strainEnergyFunction}{\firstInvariant, \secondInvariant} \left(\overline{\mathbf{F}}\right) + p\left[\det \mathbf{F}-1\right]\right] \mathbf{I}-\mathbf{F}^\text{T}\cdot \left[ \frac{\partial \interpolantVarIso{\strainEnergyFunction}{\firstInvariant, \secondInvariant} \left(\overline{\mathbf{F}}\right)}{\partial\mathbf{F}} + p \mathbf{F}^{-\text{T}} \right].
\end{equation}

\begin{remark}
The energy-momentum representation in Equation~\ref{eq:Eshelby_stress_F} allows to calculate the Eshelby stress via post processing based on the solution of the spatial motion problem.
\end{remark}

The material motion problem can be discretized via FE to obtain configurational forces at the nodes of the discretization by elements in ${\mathcal{B}_\text{e}} \in \bodyMat$ \cite{Steinmann2001}. The configurational force at global node $A$ can be calculated using the material gradient of the global node $A$ basis function ($N^A$)\footnote{The basis functions $N^A$ used to compute nodal configurational forces are equal to the function space used for the Galerkin discretization of the forward problem, i.e., linear Lagrange polynomials.}, the Eshelby stress tensor ($\boldsymbol \Sigma$), and the FE assembly operator $\opA_{e=1}^{n_\text{el}}$ over all elements
\begin{equation}\label{eq:CForce_node}
\mathbf{F}^A_\text{CNF} = \opA_{e=1}^{n_\text{el}} \int_{\mathcal{B}_\text{e}} \boldsymbol \Sigma \cdot \nabla_0 N^A \text{d}\text{V}.
\end{equation}

To achieve a better approximation of the $J$-integral the spurious configurational forces in a region surrounding the crack tip are added up to the crack tip physical configurational forces (see, e.g., \cite{Denzer2003,Moreno-Mateos2024b}). To that end, a cylindrical region of radius \qty{0.1}{\milli \meter} is defined to add the spurius forces that lay inside the cylinder close to the crack tip.

For more details on the motivation and implementation of the Configurational Force Method at finite strains, the reader may consult the dedicated work by the authors, \cite{Moreno-Mateos2024b}.

\subsection{Results}
\begin{figure}[H]
\centering
\includegraphics[width=0.9\textwidth]{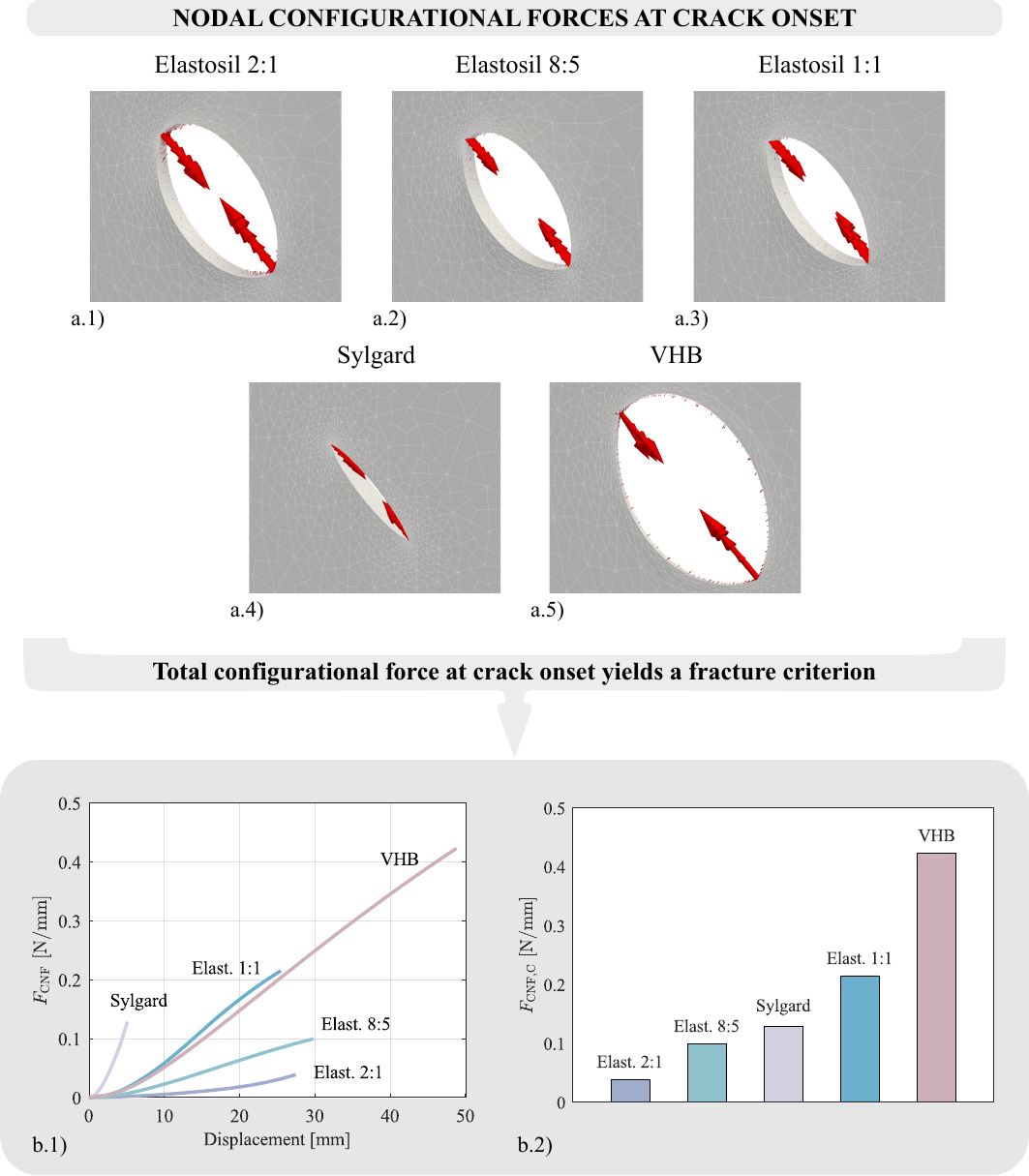}
\caption{\textbf{Results for data-adaptive configurational forces \textit{ante} crack onset and fracture criterion based on their critical values.} (a) Nodal configurational forces at the crack tip and spurious configurational forces at the crack tip vicinity are depicted for Elastosil 2:1, Elastosil 8:5, Elastosil 1:1, Sylgard, and VHB at fracture onset. The loading stage selected in the virtual experiment (computation) is exactly the one corresponding to the displacement at crack onset (average values in Table~\ref{tab:displacement_crack_onset}). Subsequently, (b.1) the nodal forces are added up to a crack tip total configurational force that is depicted as a function of the loading---displacement of the clamps--- in the virtual experiment up to fracture onset. (b.2) The critical values of the configurational force for fracture initiation are depicted in a barplot, with the smallest value for Elastosil 2:1 and the largest value for VHB.}
\label{fig:FigCFs}
\end{figure}

The data-adaptive hyperelastic energy functions identified in the previous section are now incorporated into the Configurational Force Method. Using these functions, both nodal and total configurational forces acting at the crack tip are computed at discrete loading steps throughout the deformation process---prior to crack initiation. The analysis is carried out up to the point at which fracture propagation begins, as determined experimentally from the average displacement at crack onset (see Table~\ref{tab:displacement_crack_onset}). The results, shown in Figure~\ref{fig:FigCFs}, include: (a) nodal configurational forces at crack onset, (b) evolution of the total configurational force ($\mathbf{F}_{\mathrm{CNF}}$) with clamp displacement, and (c) critical configurational force values at crack onset ($\mathbf{F}_{\mathrm{CNF,C}}$).

A first inspection of the nodal configurational forces at crack onset (Figure~\ref{fig:FigCFs}.a) shows strong localization of the forces at the nodes at the crack tip edge (physical configurational forces), alongside the presence of minor spurious forces in the vicinity of the tip. The spurious forces are of negligible magnitude. For visual clarity, the arrows depicting these forces are not uniformly scaled across materials and should not be used for quantitative comparison. Nodal configurational forces are, however, a consequence of the FE spatial discretization and, consequently, depend on the FE mesh.

A key quantity of interest is the total crack tip configurational force, ideally expressed per unit thickness, obtained by summing the nodal configurational forces. The total crack tip configurational force provides a direct approximation of the classical contour $J$-integral. As discussed by the authors in \cite{Moreno-Mateos2024b}, the configurational traction forces integrated along a contour enclosing the crack tip---commonly referred to as a \textit{Pacman-shaped} domain---are in equilibrium with the total configurational force acting directly at the crack tip. This balance of configurational forces underpins the theoretical equivalence between the $J$-integral and the Configurational Force Method. Accurate estimation of the $J$-integral under the biaxial loading conditions considered here critically depends on capturing the true constitutive response of the soft elastomers. This is effectively achieved through our framework for data-adaptive hyperelastic energy functions, enabling reliable and physically meaningful configurational force predictions.

The evolution of the $J$-integral, i.e., total configurational force during loading of the samples in the virtual experiments, is closely governed by the non-linear constitutive behavior of each soft elastomer. Quantitative comparability is provided by plotting the magnitude of total configurational force $F_{\mathrm{CNF}}$ against grip displacement in Figure~\ref{fig:FigCFs}.b. This total force is normalized by the initial sample thickness (\qty{0.5}{\milli\meter} for VHB, and \qty{2.0}{\milli\meter} for all other materials). Among the materials tested, Sylgard exhibits the steepest increase in $F_{\mathrm{CNF}}$, followed by Elastosil 1:1 and VHB. Elastosil 8:5 and Elastosil 2:1 show the lowest magnitudes. These trends are consistent with the initial slopes of the identified data-adaptive hyperelastic energy functions: highest for Sylgard and lowest for Elastosil 2:1. Overall, the shape of the curves is influenced by both the material's constitutive nonlinearity and the geometric nonlinearities inherent in the experimental setup.

In the spirit of a critical value of the $J$-integral or stress intensity factor in the context of the rather limited linear elastic fracture mechanics, the total crack tip configurational force at crack onset serves as a measure of the material's fracture toughness. This leads to what we term a configurational-forced-based fracture criterion, in which fracture onset (not complete rupture) is predicted when the magnitude of the total configurational force reaches a critical value, $F_\mathrm{CNF}=F_{\mathrm{CNF,C}}$. The bar plot in Figure~\ref{fig:FigCFs}.c presents the critical configurational forces $F_{\mathrm{CNF,C}}$ for the five soft elastomers investigated in this work. These values correspond to the final points of the force-displacement curves in Figure~\ref{fig:FigCFs}.b and are as follows: \qty{0.0386}{\newton \per \meter} for Elastosil 2:1, \qty{0.0994}{\newton \per \meter}  for Elastosil 8:5, \qty{0.2149}{\newton \per \meter}  for Elastosil 1:1, \qty{0.1286}{\newton \per \meter} for Sylgard, and \qty{0.4228}{\newton \per \meter} for VHB\footnote{We note that $F_{\mathrm{CNF,C}}$ for VHB yields a value approximately twice as large as that reported in \cite{Moreno-Mateos2024b}. This discrepancy is expected, as the constitutive model used in the aforementioned study was less accurate than the data-driven hyperelastic energy functions employed in the present work.}.These results may be interpreted in relation to the total work of fracture shown in Figure~\ref{fig:barplots_exp}.a. Notably, Sylgard exhibits an intermediate value of $F_{\mathrm{CNF,C}}$ but a remarkably low total work of fracture due to its limited displacement at failure, suggesting a brittle-like fracture behavior. In contrast, VHB displays the highest values of total work of fracture, displacement at failure, and $F_{\mathrm{CNF,C}}$, a hallmark of highly deformable soft elastomers.

Lastly, we write a note on the applicability of the method to the sideways-fracturing Elastosil 1:1 elastomer. Although fracture in this Elastosil variant propagates laterally, the Configurational Force Method predicts a total configurational force vector oriented in the forward-fracturing direction. This apparent inconsistency can be intuitively understood: the outcome stems from the use of an isotropic hyperelastic constitutive model and the underlying variational framework, which, sensu stricto, yields configurational forces pointing in the direction of maximal strain energy release, i.e., the direction in which a virtual crack extension (configurational change) would most efficiently reduce the total energy of the system. As discussed in \cite{Moreno-Mateos2024n}, the observed sideways fracture behavior can be explained through deformation-induced fracture anisotropy in elastomers modeled with standard (isotropic) hyperelastic strain energy functions. Within this context, and in the absence of a re-formulated, anisotropy-aware energy density, the Configurational Force Method remains capable of providing a reliable estimate of the magnitude of $\mathbf{F}_{\mathrm{CNF,C}}$, but it falls short in predicting its direction. This limitation arises because the method, as implemented here, does not account for anisotropic fracture resistance—in particular, the fact that the fracture energy is lower in the lateral direction than in the forward direction. A potential remedy here will involve microstructure-informed constitutive models that incorporate complex polymer chain-level mechanisms and their influence on fracture behavior. Readers interested in this direction may consult relevant literature, including \cite{Stumpf2010,Behnke2018,Mulderrig2021,Arunachala2024,Mulderrig2025}.

\section{Discussion and outlook}
\subsection{Discussion}
In the scope of this article, we have shown that the material characterization under multiaxial loading conditions plays a central role to accurately capture the constitutive behavior of soft elastomers. In turn, our data-adaptive framework for hyperelastic energy functions proves versatility to capture biaxial experiments for different soft constitutive behaviors. This is paramount because accurately capturing the constitutive behavior of soft elastomers is key for a reliable application of the Configurational Force Method to soft solids. 

Data-adaptive configurational forces for fracture require accurate constitutive models. Our framework combines reaction force data with full-field displacement measurements to calibrate hyperelastic energy functions. Displacement fields are extracted from equi-biaxial tests on five distinct soft elastomers---all with Young's modulus below \qty{300}{\kilo \pascal}---and featuring a centrally placed pre-cut that serves as strain concentrator and fracture initiator. The results reveal a spectrum of constitutive and fracture behaviors from brittle-like to highly deformable responses. Unlike pure VFM-based methods, which typically require full-field data across the entire surface, our FEMU-based optimization strategy remains unbiased even when the DIC data is incomplete. Notably, we find that the data-adaptive optimization of the hyperelastic energy functions based on biaxial test data primarily activates the first invariant strain energy function $\interpolantVarIso{\strainEnergyFunction}{\firstInvariant}$. 

A fracture criterion is established based on a critical value of the crack tip configurational force $F_\text{CNF,C}$. To capture the nonlinear, material-specific response of soft solids, we embed data-adaptive hyperelastic energy functions into the Configurational Force Method, enabling the computation of what we term \textit{data-adaptive configurational forces for fracture}. A dedicated computational testbed replicates the biaxial experiments and leverages finite element solutions of the forward boundary value problem to evaluate the Eshelby stress tensor and associated nodal configurational forces. In the spirit of a fracture toughness material parameter, the crack tip configurational force at crack onset is a computationally efficient estimator of the $J$-integral. This approach captures the intricate constitutive behavior of soft materials under large geometrical nonlinearities, thereby offering a robust and efficient framework for fracture fracture in highly deformable solids.

\subsection{Outlook}
Several experimental directions remain open for biaxial characterization of soft materials. Future studies could explore, among others, the effects of strain rate and deviations from equi-biaxial loading. In addition, promising opportunities lie in the multiaxial characterization of macroscale heterogeneous soft materials \cite{Li2020a,Moreno-Mateos2024n}. In such systems, different fracture modes, such as sideways fracturing, could be studied together with micromechanical models that capture the intrinsic fracture mechanisms of soft polymers (see, e.g., \cite{Mulderrig2021}).

A particularly promising extension of our framework for data-driven hyperelastic energy functions is its application to rate-dependent (viscoelastic) constitutive behavior. This may be achieved via a multiplicative decomposition of the deformation gradient into elastic and viscous parts, building upon a finite viscoelasticity model, such as the one proposed in \cite{reese_theory_1998}. 

Although our FEMU-based optimization method has demonstrated efficiency and robustness, a promising future direction lies in integrating of our data-adaptive hyperelastic energy functions with Physics-Informed Neural Networks. Recent work has shown that hyperelastic constitutive behavior can be represented using feed-forward neural networks \cite{hamel_calibrating_2023}. This approach may eliminate the need for finite element forward simulations, making it an interesting benchmark for comparison.

Furthermore, the insights gained from our data-adaptive configurational force framework open new avenues for reverse engineering strategies that tailor the constitutive and fracture properties of soft materials through informed design and manufacturing parameters. In particular, neural network–based surrogate models hold promise for efficiently mapping observed mechanical behavior---such as hyperelasticity and fracture toughness---back to material formulation or processing variables. In this reversed workflow, learned parameters related to fracture serve as inputs, while outputs correspond to fabrication pathways related to, e.g., elastomer synthesis. In this context, our framework may provide a systematic and high-fidelity route to generate the curated datasets required to train such models, thereby enabling data-driven material design rooted in fracture mechanics.

Lastly, a point of interest is the extension to our framework to the \textit{post} fracture onset regime. In this context, the use of configurational forces together with a fracture dissipation inequality may open new interesting routes to enable predictive modeling beyond crack initiation, paving the way toward a more comprehensive understanding of failure in highly deformable materials.

\newpage

\section*{Data Availability}
The experimental datasets are available in \url{https://doi.org/10.5281/zenodo.15187640}.

\section*{Acknowledgments}
Miguel Angel Moreno-Mateos, Simon Wiesheier, and Paul Steinmann acknowledge support from the European Research Council (ERC) under the Horizon Europe program (Grant-No. 101052785, project: SoftFrac). Funded by the European Union. Views and opinions expressed are however those of the author(s) only and do not necessarily reflect those of the European Union or the European Research Council Executive Agency. Neither the European Union nor the granting authority can be held responsible for them. Mokarram Hossain acknowledges support from the Engineering and Physical Sciences Research Council (EPSRC) under the grant (EP/Z535710/1).
\begin{figure}[H]
\includegraphics[width=0.3\textwidth]{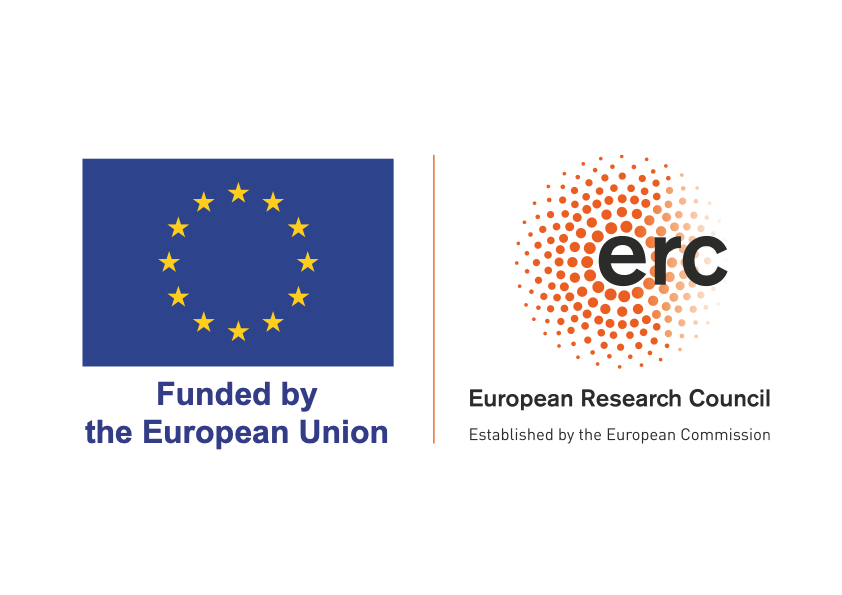}
\end{figure}

\newpage

\appendix
\section{Additional results for strain fields}\label{sec:Additional_Strain_Fields}

\begin{figure}[H]
\centering
\includegraphics[width=0.7\textwidth]{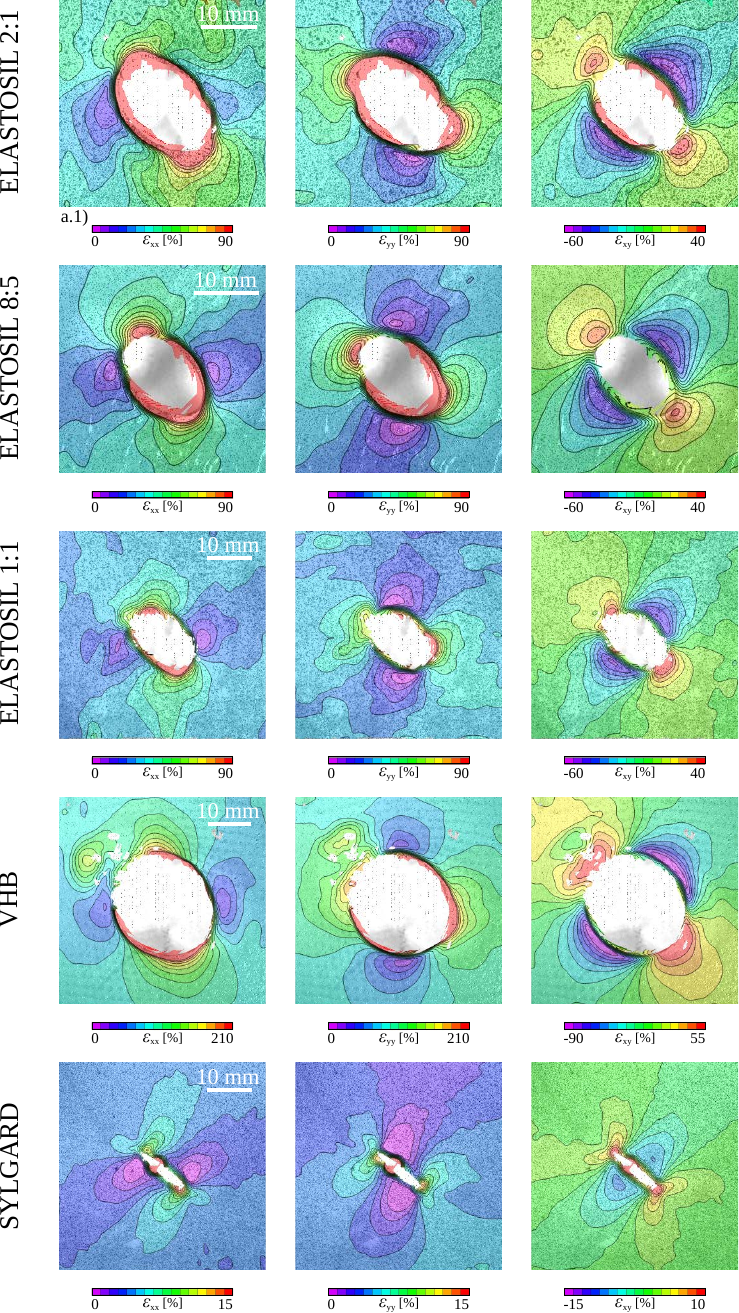}
\caption{\textbf{Strain fields at onset of crack growth.} The fields are engineering strain fields computed from the Lagrange strain tensor. 
The fields correspond to the loading stages described in Figure~\ref{fig:Force-Disp}.a.1-5 and in Table~\ref{tab:displacement_crack_onset}. The fields correspond for the second repetition (out of the four repetitions performed for the same type of sample and same test conditions).  Note that $x$ and $y$ denote the horizontal and  vertical directions, respectively.}
\label{fig:FigDICStrains2}
\end{figure}

\begin{figure}[H]
\centering
\includegraphics[width=0.7\textwidth]{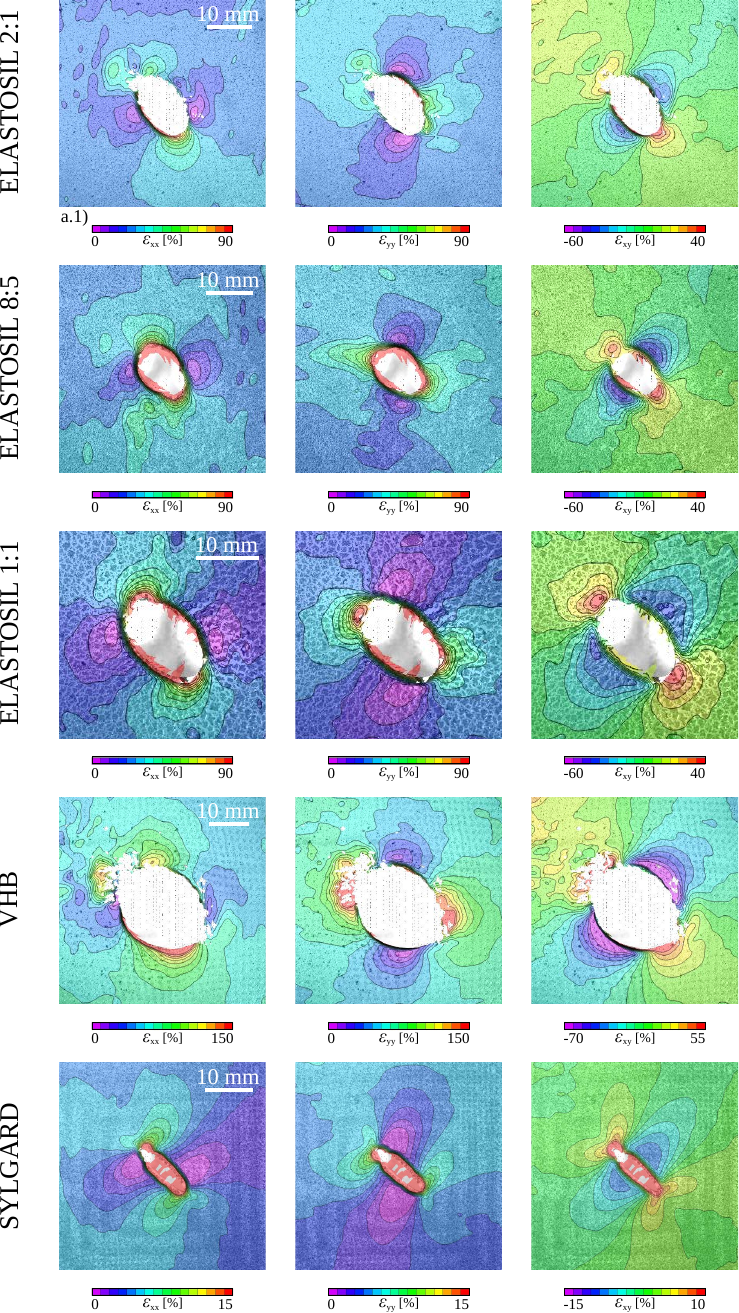}
\caption{\textbf{Strain fields at onset of crack growth.} The fields are engineering strain fields computed from the Lagrange strain tensor. 
The fields correspond to the loading stages described in Figure~\ref{fig:Force-Disp}.a.1-5 and in Table~\ref{tab:displacement_crack_onset}. The fields correspond for the third repetition (out of the four repetitions performed for the same type of sample and same test conditions).  Note that $x$ and $y$ denote the horizontal and  vertical directions, respectively.}
\label{fig:FigDICStrains3}
\end{figure}

\begin{figure}[H]
\centering
\includegraphics[width=0.7\textwidth]{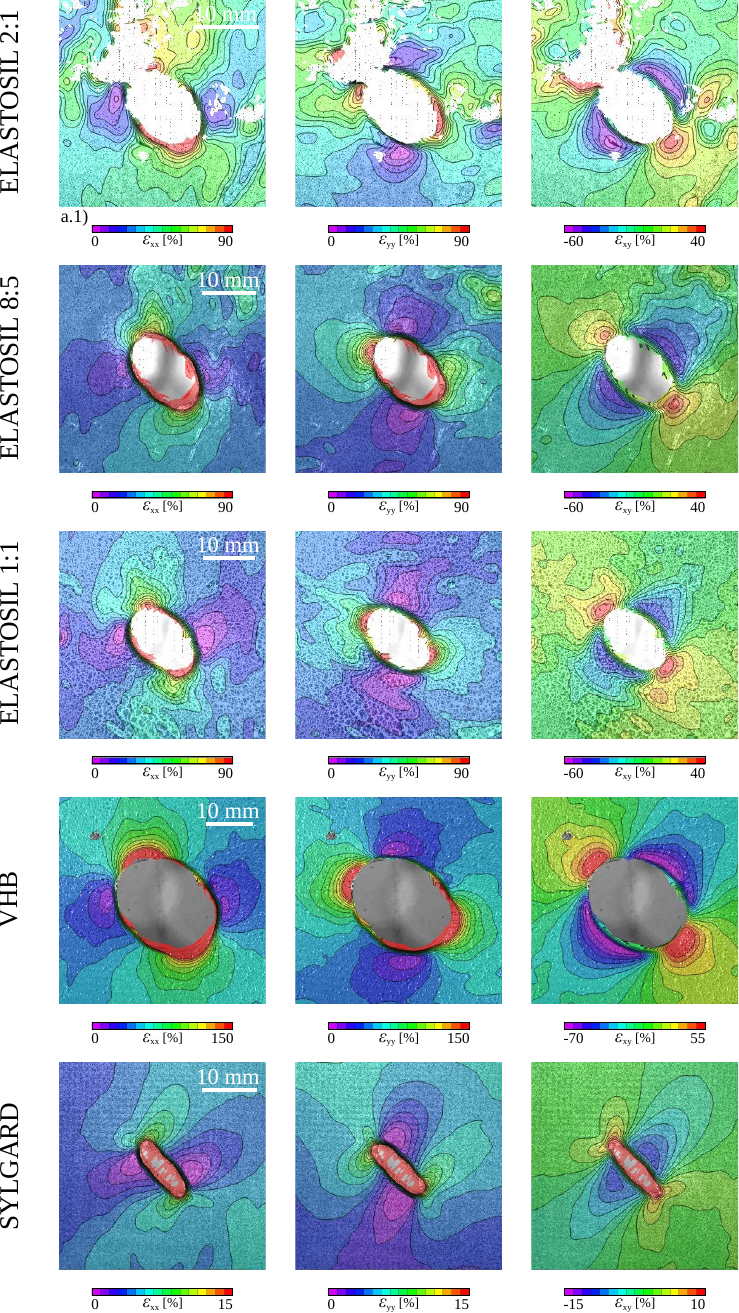}
\caption{\textbf{Strain fields at onset of crack growth.} The fields are engineering strain fields computed from the Lagrange strain tensor. 
The fields correspond to the loading stages described in Figure~\ref{fig:Force-Disp}.a.1-5 and in Table~\ref{tab:displacement_crack_onset}. The fields correspond for the fourth repetition (out of the four repetitions performed for the same type of sample and same test conditions).  Note that $x$ and $y$ denote the horizontal and  vertical directions, respectively.}
\label{fig:FigDICStrains4}
\end{figure}

\section{Kernel Density Estimation}
\label{sec:SecKernel}

\begin{figure}[H]
\centering
\includegraphics[width=0.7\textwidth]{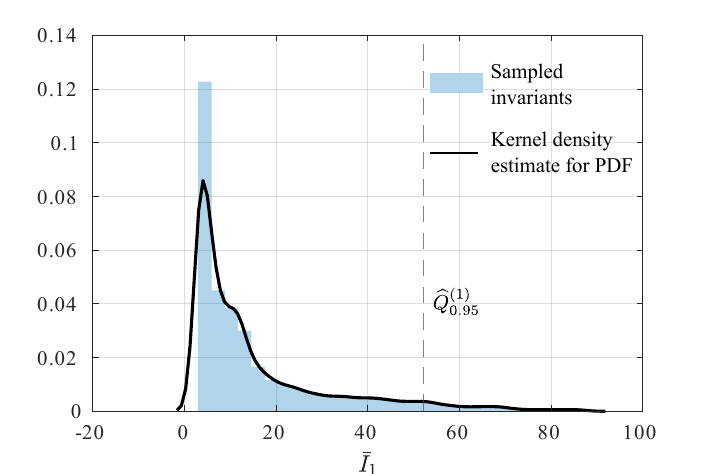}
\caption{\textbf{Kernel Density Estimation (KDE) employed to obtain a differentiable representation of the probability density function (PDF) for the sampled invariants $\firstInvariant, \secondInvariant$.} }
\label{fig:FigI1Density}
\end{figure}

At each optimization iteration, the sampled invariants ($\firstInvariant, \secondInvariant$) at the quadrature points of the FE-mesh are transferred into a histogram. Figure~\ref{fig:FigI1Density} illustrates mocked $\firstInvariant$ values obtained from a forward solve. While a histogram plot groups these samples into discrete bins and yields a discrete probability density function (PDF), the resulting representation is not continuously differentiable. This is problematic because the gradient-based optimizer \textit{fmincon} requires the objective function to be continuously differentiable with respect to all optimization variables. For example, computing the 95-quantile $\quantile{0.95}{1}$ directly from the histogram would violate this differentiability assumption. 

To address this, we apply kernel density estimation (KDE), which produces a smooth and continuously differentiable approximation of the PDF. In this approach, each sampled value contributes a smooth kernel function, and the overall PDF is obtained by summing these kernels. 

The required 95-quantile $\quantile{0.95}{1}$, as used in Equation~\eqref{eq:EqLoss}, is extracted by evaluating the inverse cumulative density function (iCDF) of the KDE
\begin{align}
    \quantile{0.95}{1} =: \iCDF{\firstInvariant}(0.95).
\end{align}

Using a Gaussian kernel, the PDF and the corresponding CDF are given by
\begin{alignat}{2}
\text{PDF:} \quad \pdf{\firstInvariant}(x) 
&= \frac{1}{nh} \sum\limits_{i=1}^n K\left(\frac{x - x_i}{h}\right) \quad 
&&K(x) = \frac{1}{\sqrt{2\pi}} \exp\left(-\frac{1}{2}x^2\right) \\[0.5em]
\text{CDF:} \quad \cddf{\firstInvariant}(x) 
&= \frac{1}{n} \sum\limits_{i=1}^n G\left( \frac{x - x_i}{h} \right) \quad 
&&G(x) = 0.5\left[1 + \text{erf}\left(\frac{x}{\sqrt{2}}\right) \right],
\end{alignat}
\noindent with $n$ the number of samples.

The bandwidth $h$ is defined as
\begin{align}
    h = \left[\frac{4}{3n}\right]^{1/5} \sigma,
\end{align}
where $\sigma$ is the standard deviation of the samples. This choice is optimal for normally distributed data and provides a sufficiently accurate approximation of the sampled invariant distribution for our purposes \cite{atkinson_oxford_nodate}. We aggregated all samples at the quadrature points in each load step of the forward solve. A more efficient sampling strategy (e.g., sampling invariants only at specific geometric locations and load steps) is left for future work. 
The above steps are identically applied to the second invariant $\secondInvariant$.

\section{Sampled invariants in the FE forward boundary value problem}
\label{sec:SecInvariants}

\begin{figure}[H]
\centering
\includegraphics[width=0.6\textwidth]{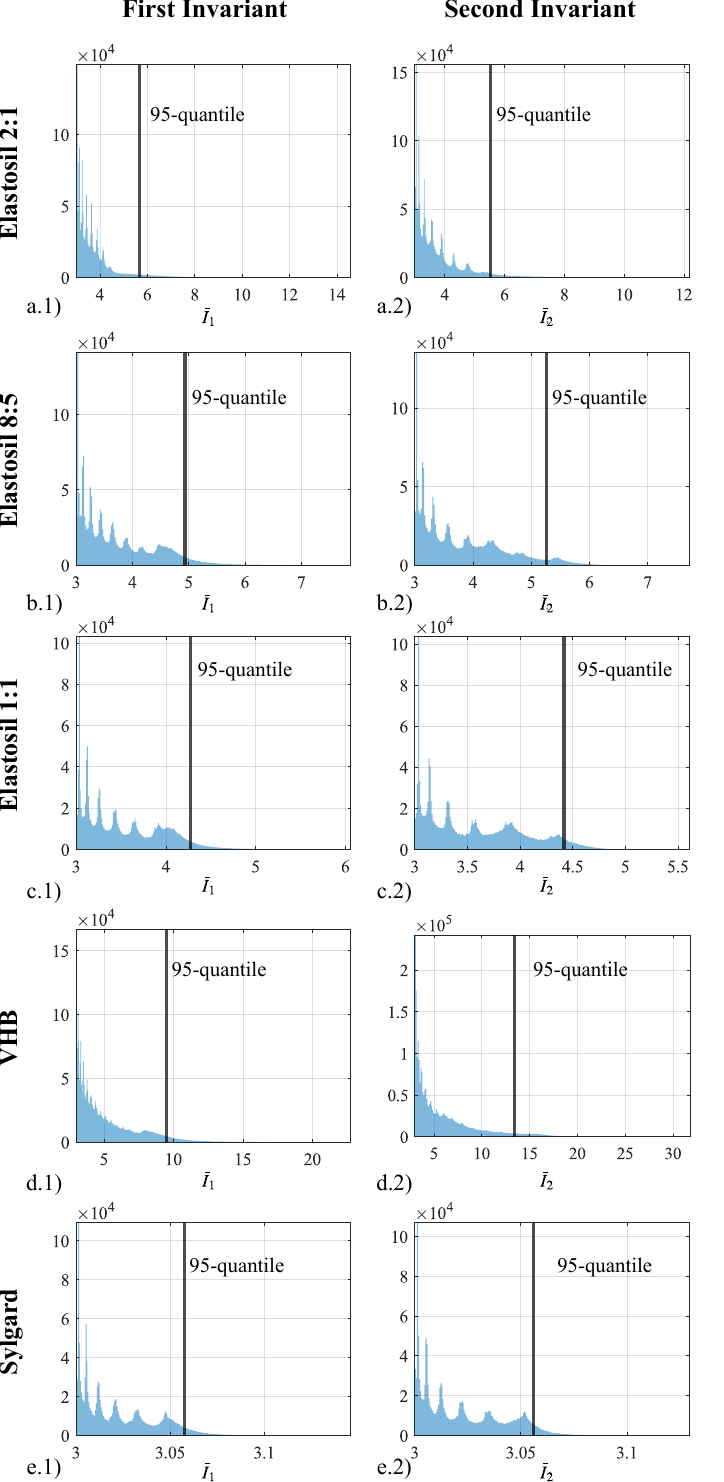}
\caption{\textbf{Sampled invariants predicted by the FE forward simulation using the identified data-adaptive hyperelastic strain energy functions.} The vertical line marks the 95th-percentile of the invariants.}
\label{fig:FigInvariants}
\end{figure}

The histograms in Figure~\ref{fig:FigInvariants} show the distributions of the invariants $\firstInvariant$ and $\secondInvariant$ as predicted by the forward simulations for each material’s identified strain energy function. The histograms are constructed from the invariant values evaluated at all quadrature points of the finite element mesh, aggregated across all simulation load steps. The vertical lines mark the 95th percentile of each distribution.

The overall range and shape of the $\firstInvariant$ and $\secondInvariant$ distributions are similar for the three Elastosil variants and for Sylgard. In contrast, the VHB material exhibits a noticeably wider spread between the two invariants, as reflected in the respective 95th-percentile values, which are \qty{9.52}{} for $\firstInvariant$ and \qty{13.44}{} for $\secondInvariant$. This behavior can be qualitatively understood by considering the simplified case of equibiaxial tension, where the invariants take the form \cite{steinmann_hyperelastic_2012}:
\begin{align}
\firstInvariant &= \lambda^{-4} + 2\lambda^2, \\
\secondInvariant &= 2\lambda^{-2} + 2\lambda^4.
\end{align}
These expressions illustrate that the difference between $\firstInvariant$ and $\secondInvariant$ increases with the applied deformation $\lambda$. However, in our biaxial experiments, the applied deformation (less than \qty{100}{\%} engineering strain) is not large enough to induce a pronounced separation between the two invariants. Among all materials, Sylgard exhibits the narrowest range of sampled invariants, staying below \qty{3.1}{}. The visible peaks in the histograms correspond to the invariant values sampled at discrete simulation load steps.

A key quantity extracted from the histograms is the 95th-percentile value, which we approximate using kernel density estimation (cf. \ref{sec:SecKernel}). This estimate helps align the interpolation domain of the strain energy function with the range of actually sampled invariants. Comparing the rightmost interpolation point in Figures~\ref{fig:FigElastosil21}–\ref{fig:FigVHB}a with the 95th-percentile values in Figure~\ref{fig:FigInvariants} indicates good agreement. This demonstrates that the discretized invariant space of our data-driven hyperelastic energy functions is neither too narrow nor unnecessarily wide relative to the actual sampled invariants triggered by the data.

\newpage
\bibliographystyle{naturemag}
\addcontentsline{toc}{section}{References}

\end{document}